\documentclass[aps,prl,twocolumn,superscriptaddress,showpacs]{revtex4-1}

\usepackage{amsmath,amssymb,graphics,epsfig,epstopdf,color,verbatim,ulem,braket,tabularx,float}
\usepackage[colorlinks,linkcolor=blue,citecolor=blue,urlcolor=blue]{hyperref}

\graphicspath{{figures/}}
\begin{document}

\title{SrPd, a candidate material with extremely large magnetoresistance}
\author{Xiao-Qin Lu}
\affiliation{Department of Physics and Beijing Key Laboratory of Opto-electronic Functional Materials $\&$ Micro-nano Devices, Renmin University of China, Beijing 100872, China}

\author{Peng-Jie Guo}
\affiliation{Songshan Lake Materials Laboratory, Dongguan, Guangdong 523808, China}

\author{Jian-Feng Zhang}
\affiliation{Department of Physics and Beijing Key Laboratory of Opto-electronic Functional Materials $\&$ Micro-nano Devices, Renmin University of China, Beijing 100872, China}

\author{Kai Liu}\email{kliu@ruc.edu.cn}
\affiliation{Department of Physics and Beijing Key Laboratory of Opto-electronic Functional Materials $\&$ Micro-nano Devices, Renmin University of China, Beijing 100872, China}

\author{Zhong-Yi Lu}\email{zlu@ruc.edu.cn}
\affiliation{Department of Physics and Beijing Key Laboratory of Opto-electronic Functional Materials $\&$ Micro-nano Devices, Renmin University of China, Beijing 100872, China}

\begin{abstract}

The extremely large magnetoresistance (XMR) effect in nonmagnetic semimetals have attracted intensive attention recently. Here we propose an XMR candidate material SrPd based on first-principles electronic structure calculations in combination with a semi-classical model. The calculated carrier densities in SrPd indicate that there is a  good electron-hole compensation, while the calculated intrinsic carrier mobilities are as high as 10$^5$ cm$^2$V$^{-1}$s$^{-1}$. There are only two doubly degenerate bands crossing the Fermi level for SrPd, thus a semi-classical two-band model is available for describing its transport properties. Accordingly, the magnetoresistance of SrPd under a magnetic field of $4$ Tesla is predicted to reach ${10^5} \%$ at low temperature. Furthermore, the calculated topological invariant indicates that SrPd is topologically trivial. Our theoretical studies suggest that SrPd can serve as an ideal platform to examine the charge compensation mechanism of the XMR effect.

\end{abstract}


\date{\today}
\maketitle

\section{Introduction}

The magnetoresistance (MR) is defined as the change in electrical resistance with applied magnetic field, which has promising applications in magnetic sensors, hard drives, and other electronic devices~\cite{lenz1990review, daughton1999gmr, reig2009magnetic}. The MR phenomena have been intensively studied in magnetic compounds, which include anisotropic magnetoresistance in ferromagnetic metals~\cite{mcguire1975anisotropic}, giant magnetoresistance in magnetic thin-film superlattices~\cite{PhysRevLett.61.2472}, colossal magnetoresistance in rare-earth doped oxides~\cite{dagotto2001colossal, salamon2001physics}, and tunneling magnetoresistance in ferromagnetic tunnelling junctions~\cite{moodera1995large, parkin2004giant}. On the other hand, the conventional MR in many nonmagnetic metals originating from the Lorentz force only shows a small value of a few percentage~\cite{pippard1989magnetoresistance, stohr2006and}. Recently, several studies revealed an extremely large magnetoresistance (XMR) as high as ${10^5}\%$ to ${10^7}\%$ in nonmagnetic materials with distinct topological properties, such as topologically trivial semimetals LaSb~\cite{tafti2016resistivity, LaSbShanShan2016, PhysRevB.93.235142, PhysRevLett.117.127204} and YSb~\cite{YSbQiaoHe2017, PhysRevLett.117.267201}, Dirac semimetals Cd$_3$As$_2$~\cite{liang2015ultrahigh} and NbSb$_2$~\cite{wang2014anisotropic}, and Weyl semimetals WTe$_2$~\cite{ali2014large, LvWTe2, 2015PRL.115.WTe2SOC}, and TaAs~\cite{PhysRevX.5.031023}. The mechanism underlying such an XMR effect is still an ongoing research project~\cite{ZhangSN-arxiv}.

The exploration of the XMR mechanism has motivated many experimental and theoretical studies. One explanation is the back-scattering protection mechanism. Due to existence of topological protection, the electrons' back scattering is suppressed at zero magnetic field~\cite{ali2014large, LvWTe2, liang2015ultrahigh, shekhar2015extremely, PhysRevB.92.205134} but opens under magnetic field. Another scenario is the electron-hole compensation mechanism, which is based on the semi-classical two-band model~\cite{ali2014large, PhysRevLett.113.216601}. The XMR phenomena in type-II Weyl semimetals WTe$_2$~\cite{ali2014large, LvWTe2, 2015PRL.115.WTe2SOC} and WP$_2$~\cite{2017WP2-ncomn, PhysRevB.96.121107, PhysRevB.96.121108} were explained by both of the above scenarios. Recently, it was found that topologically trivial semimetal LaSb and nontrivial semimetal LaBi both show nonsaturating XMR effect~\cite{tafti2016resistivity, LaSbShanShan2016, PhysRevB.93.235142, PhysRevLett.117.127204, PhysRevB.95.115140}. The common feature between these two rare-earth pnictide materials is the electron-hole compensation, indicating a key role that it may play in the XMR effect. Nevertheless, these XMR materials are actually not exact two-band systems since there are three doubly degenerate bands across the Fermi level. To further ascertain the charge compensation mechanism, it is important to find more materials with XMR effect but trivial electronic structure.

In this paper, we have theoretically studied the electronic structures, the topological properties, and the magnetoresistance properties of SrPd and BaPd. The electron-hole compensation and high carrier mobilities in SrPd have been derived from first-principles electronic structure calculations. Moreover, SrPd is found to be a trivial semimetal. As there are only two bands crossing the Fermi level, SrPd is predicted to be a good candidate of XMR material that can be well described by a semiclassical two-band model.

\section{Calculation method}

The first-principles electronic structure calculations were carried out by using the projector augmented wave (PAW) method~\cite{PhysRevB.50.17953, PhysRevB.59.1758} as implemented in the VASP package~\cite{PhysRevB.47.558, KRESSE199615, PhysRevB.54.11169}. The generalized gradient approximation (GGA) of Perdew-Burke-Ernzerhof (PBE) type~\cite{PhysRevLett.77.3865} was adopted for the exchange-correlation functional. The kinetic energy cutoff of the plane-wave basis was set to be 350 eV. A 16$ \times $16$ \times $20 $k$-point mesh was used for the Brillouin zone (BZ) sampling of primitive cell. The Gaussian smearing method with a width of 0.01 eV was adopted for the Fermi surface broadening. Both cell parameters and internal atomic positions were fully relaxed until the forces on all atoms were smaller than 0.01 eV/{\AA}. To compare with the PBE results at GGA level, the strongly constrained and appropriately normed (SCAN)~\cite{PhysRevLett.115.036402} semilocal density functional at meta-GGA level of the Jacob's ladder~\cite{Perdew_J.Chem.Phys2015} was also used. Once the equilibrium structures were obtained, the spin-orbit coupling (SOC) effect was included in the electronic structure calculations. To analyze the Fermi surface, the maximally localized Wannier functions' (MLWF) method~\cite{PhysRevB.56.12847, PhysRevB.65.035109, MOSTOFI2008685} was employed. The topological invariant Z2~\cite{PhysRevLett.95.146802} was calculated based on the Wilson loops~\cite{PhysRevB.84.075119} (equivalent to Wannier charge centers~\cite{PhysRevB.83.235401}) method by using the WannierTools package~\cite{WU2018405}.

The intrinsic carrier mobilities were studied based on the semi-classical Boltzmann transport equation and self-energy relaxation time approximation, as implemented in EPW package~\cite{epw20102140,PhysRevB.76.165108} that interfaces with Quantum ESPRESSO (QE) software~\cite{Giannozzi_2009}. The specific formulae are as follows. By ignoring the impurity scattering, we can obtain the intrinsic electron mobilities as~\cite{PhysRevB.97.121201,2015Zhou14777}:
  \begin{equation}
\mu _{e,\alpha\beta}  = \frac{{-e}}{{{n_e\Omega}}}\sum_{n\in CB}{\int\frac{d\textbf{k}}{\Omega_{BZ}}\frac{\partial f^0_{n\textbf{k}}}{\partial \epsilon_{n\textbf{k}}}v_{n\textbf{k},\alpha}v_{n\textbf{k},\beta}\tau^0_{n\textbf{k}}},
\end{equation}
where $n_e$ is the electron-type carrier density, $\Omega$ and $\Omega_{BZ}$ are respectively the volumes of unit cell and first BZ, $f^0_{n\textbf{k}}$ is the Fermi-Dirac distribution of electrons, $\epsilon_{n\textbf{k}}$ is the electronic eigenvalue, $v_{n\textbf{k},\alpha}=\frac{\partial\epsilon_{n\textbf{k}}}{\hbar\partial{k_\alpha}} $ is the band velocity, and $\alpha$ ($\beta$) means the direction of electric current. The electron-phonon relaxation time $\tau^0_{n\textbf{k}}$ has the following form,
  \begin{eqnarray}
\frac{1}{\tau^0_{n\textbf{k}}} =&&
                        \frac{2\pi}{\hbar}\sum_{m\nu}\int\frac{d\textbf{q}}{\Omega_{BZ}}|g_{mn\nu(\textbf{k},\textbf{q})}|^2\times [(1-f^0_{mk+\textbf{q}}+n_{\textbf{q}\nu})\times \nonumber\\
                       && \delta(\epsilon_{n\textbf{k}}-\epsilon_{m\textbf{k}+\textbf{q}}-\hbar\omega_{\textbf{q}\nu})+(f^0_{m\textbf{k}+\textbf{q}}+n_{\textbf{q}\nu})\times \nonumber\\
                       &&\delta(\epsilon_{n\textbf{k}}-\epsilon_{m\textbf{k}+\textbf{q}}+\hbar\omega_{\textbf{q}\nu})],
\end{eqnarray}
where $g_{mn\nu(\textbf{k},\textbf{q})}=<u_{m\textbf{k}+\textbf{q}}|\Delta_{\textbf{q}\nu}V_{scf}|u_{n\textbf{k}}>$ is the electron-phonon coupling matrix element, $\Delta_{\textbf{q}\nu}V_{scf}$ is the variation of Kohn-Sham potential, and $n_{\textbf{q}\nu}$ is the Bose-Einstein distribution of phonons. In the expression of relaxation time, the first and second terms describe the emission and absorption of phonons, respectively. In the calculation,  we used the 6$ \times $6$ \times $9 $k$-mesh and 2$ \times $2$ \times $3 $q$-mesh as coarse grids and then interpolated to the 24$ \times $24$ \times $36 $k$-mesh and 12$ \times $12$ \times $18 $q$-mesh as dense grids for the BZ sampling instead of integrations in Eq. (1) and (2).

\section{Results and analysis}

\begin{figure}[!t]
  \centering
  \includegraphics[width=0.98\columnwidth]{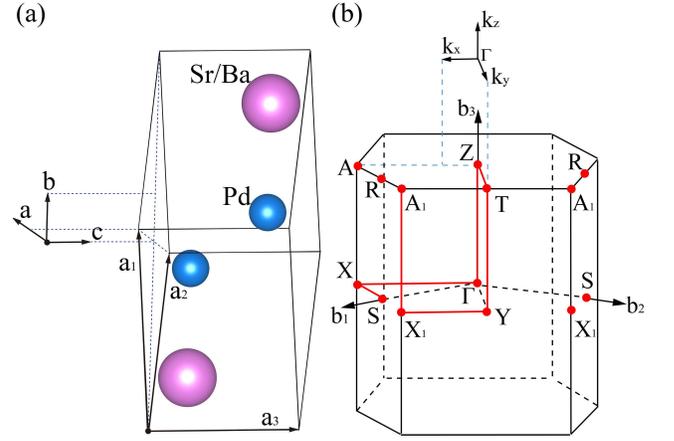}
  \caption{\label{fig:Latt}(Color online) (a) Crystal structure of SrPd and BaPd with the $Cmcm$ symmetry. The lattice vectors \textbf{a$_1$}, \textbf{a$_2$}, and \textbf{a$_3$} of primitive cell are also labeled. (b) Bulk Brillouin zone (BZ) for primitive cell of SrPd or BaPd. The red dots represent the high-symmetry $k$ points in BZ. The reciprocal vectors $k_x$, $k_y$, $k_z$ correspond to the orthogonal $a$-axis, $b$-axis, and $c$-axis in real space, respectively.}
\end{figure}

The crystal structure of SrPd has the $Cmcm$  space group symmetry~\cite{IANDELLI19741} as shown in Fig.~\ref{fig:Latt}(a), in which the centrosymmetry can be perceived. The lattice vectors of primitive cell of SrPd are labeled as \textbf{a$_1$}, \textbf{a$_2$}, and \textbf{a$_3$}, respectively. The angle between \textbf{a$_1$} and \textbf{a$_2$} vectors is $40 ^\circ$, while \textbf{a$_3$} vector is perpendicular to both of them. Figure ~\ref{fig:Latt}(b) displays the Brillouin zone of primitive cell. The calculated lattice constants of SrPd are $a = 4.219$ {\AA}, $b = 11.309$ {\AA}, and $c = 4.526$ {\AA}, which agree well with the corresponding experimental values ($a = 4.190$ {\AA}, $b = 11.310$ {\AA}, $c = 4.520$ {\AA})~\cite{IANDELLI19741}.

\begin{figure}[!t]
  \centering
  \includegraphics[width=0.98\columnwidth]{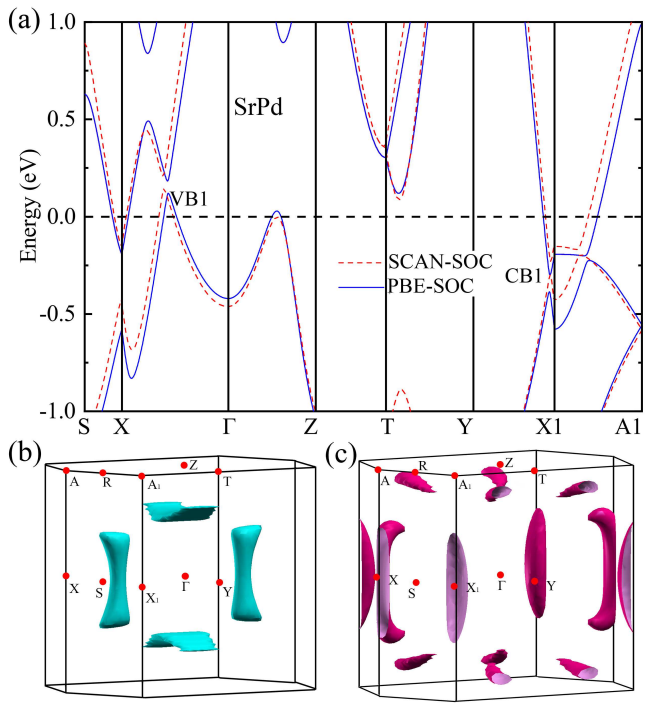}
  \caption{\label{fig:band}(Color online) (a) Band structures of SrPd along the high-symmetry paths of primitive cell calculated with the PBE functional (blue solid line) and the SCAN functional (red dashed line) including the spin-orbit coupling (SOC) effect. The valence band (VB1) and the conduction band (CB1) are marked. (b) Hole-type and (c) electron-type Fermi surface sheets of SrPd in primitive cell calculated with the PBE functional including the SOC effect. }
\end{figure}

The band structure of SrPd along the high-symmetry paths in the BZ of primitive cell is shown in Fig.~\ref{fig:band}(a). The results calculated respectively with the PBE functional at GGA level (blue solid line) and the SCAN functional at meta-GGA level (red dashed line) are both presented. We can see that the band dispersions have no essential changes for these two different functionals, thus we only concentrate on the PBE results in the following. Since SrPd is a nonmagnetic material with  centrosymmetry, its electronic bands are all doubly degenerate. In comparison with LaSb and LaBi that have three doubly degenerate bands crossing the Fermi level~\cite{PhysRevB.93.235142}, SrPd has only two doubly degenerate bands crossing the Fermi level, as shown in Fig.~\ref{fig:band}(a). Substituting Sr with Ba, BaPd has the similar characteristics as SrPd (see Fig.~\ref{fig:band2} in Appendix). Thus, both SrPd and BaPd can be well described by a semiclassical two-band model for their transport properties, as shown in the following Eq. (3).

The Fermi surface of SrPd calculated with the PBE functional including the SOC effect is also shown in Fig. ~\ref{fig:band}. There are two hole-type pockets looking like curved bread slices along the $\Gamma-Z$ direction and two other pillow-like ones along the $\Gamma-X$ direction [Fig.~\ref{fig:band}(b)]. The electron-type pockets are around the BZ boundaries [Fig.~\ref{fig:band}(c)]. From the shapes of these Fermi pockets, we can learn that SrPd exhibits a highly anisotropic property. Furthermore, the carrier densities can be obtained by calculating the volumes of these Fermi surface pockets. We have tested the 16$ \times $16$ \times $20 and 20$ \times $20$ \times $26 $k$-point meshes, which give consistent and converged results. As listed in Table ~\ref{tab:I}, the hole-type and electron-type carrier densities of SrPd are 1.21$\times$$10^{20}$ cm$^{-3}$ and 1.30$\times$$10^{20}$ cm$^{-3}$, respectively, indicating that SrPd is a semimetal with a good electron-hole compensation. Similarly, BaPd is also a charge-compensated semimetal.

The electrical resistivity of SrPd under a magnetic field can be described by the following semiclassical two-band model~\cite{solid_phys, Elec_Phon}:
\begin{equation}
\rho \left( B \right) = \frac{{\left( {{n_e}{\mu _e} + {n_h}{\mu _h}} \right) + \left( {{n_e}\mu _h^{}{\rm{ + }}{n_h}\mu _e^{}} \right)\mu _e^{}\mu _h^{}{B^2}}}{{{{\left( {{n_e}{\mu_e} + {n_h}{\mu _h}} \right)}^2}e + {{\left( {{n_e} - {n_h}} \right)}^2}\mu _e^2\mu _h^2e{B^2}}},
\end{equation}
where $n_e$ ($n_h$) is the electron-type (hole-type) carrier density, $\mu_e$ ($\mu_h$) is the electron-type (hole-type) carrier mobility, $e$ is the charge unit, and $B$ is the magnetic field perpendicular to the current direction. According to the above calculations, we learn that SrPd is a charge compensated semimetal with equal electron-type and hole-type carrier densities (Table~\ref{tab:I}). With $n={n_e}={n_h}$, the electrical resistivity can be expressed as
\begin{equation}
\rho \left( B \right) = \frac{{1 + \mu _e^{}\mu _h^{}{B^2}}}{{n\left( {{\mu _e} + {\mu _h}} \right)e}}.
\end{equation}
The magnetoresistance (MR) can thus be reduced to
 \begin{eqnarray}
 {\rm{MR}} &=& \frac{{\rho \left( B \right) - \rho \left( 0 \right)}}{{\rho \left( 0 \right)}}  \nonumber\\
 &=& \mu _e^{}\mu _h^{}{B^2} \times 100\%.
 \end{eqnarray}
From this equation, we can see that for a charge-compensated semimetal, the MR is proportional to the product of carrier mobilities and exhibits a quadratic dependence on the magnetic field.

\begin{figure}[!t]
  \centering
  \includegraphics[width=0.98\columnwidth]{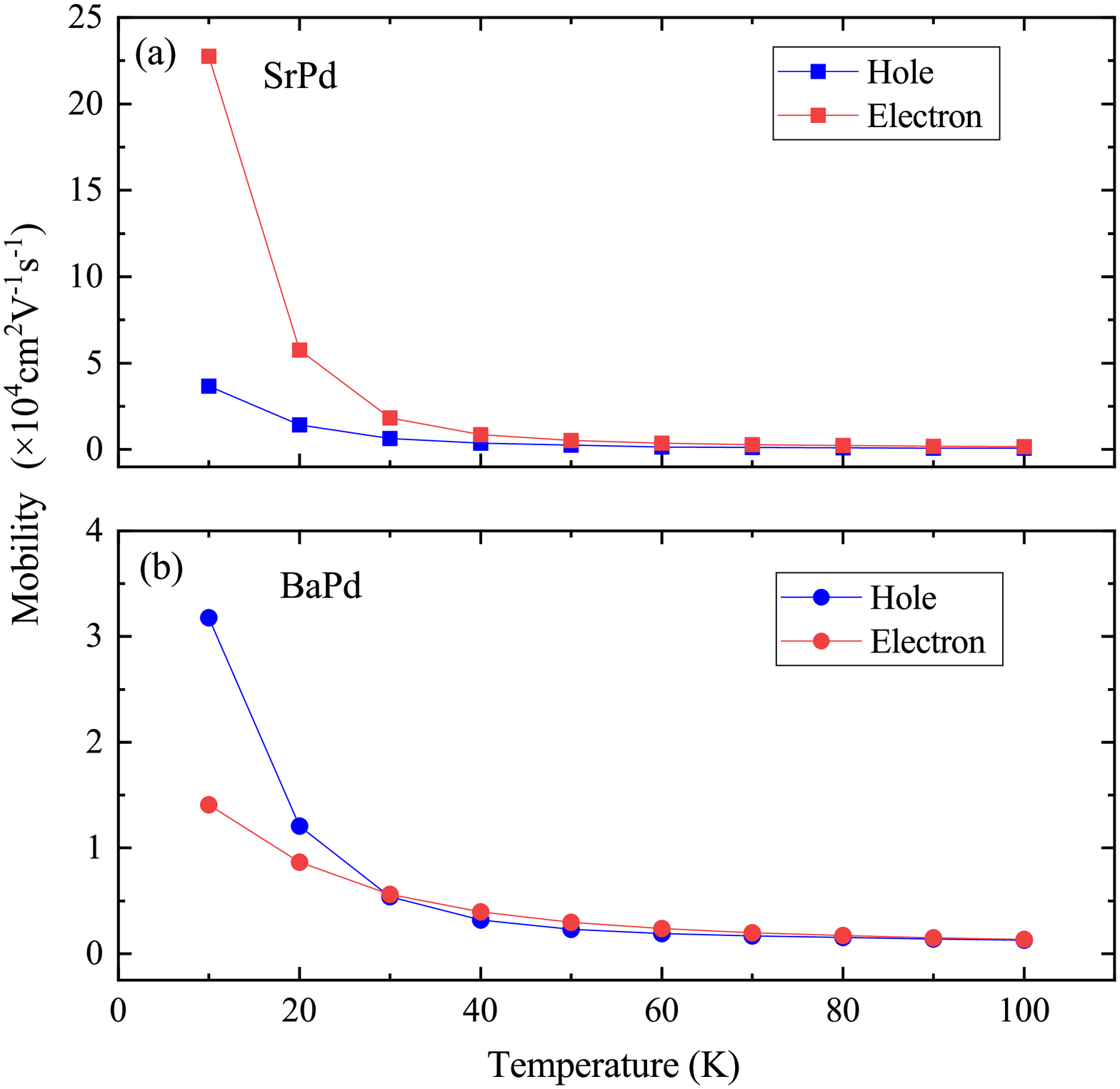}
  \caption{\label{fig:mob}(Color online) Temperature-dependent hole-type and electron-type carrier mobilities of SrPd and BaPd.}
\end{figure}


\renewcommand\arraystretch{1.5}
\begin{table}[!b]
\caption{\label{tab:I} Carrier densities $n_h$ ($n_e$) (in unit of ${10^{20}}$ cm${^{-3}}$) of SrPd and BaPd calculated with the PBE functional and the SCAN functional, respectively.}
\begin{center}
\begin{tabular*}{1.0\columnwidth}{@{\extracolsep{\fill}}ccccccc}
\hline\hline
                &    &   \multicolumn{2}{c}{SrPd}  &      &   \multicolumn{2}{c}{BaPd}  \\

 \cline{3-4} \cline{6-7}
               &     & PBE         &    SCAN          &      &   PBE         &    SCAN   \\
\hline
   $n_{h}$      &     &   1.21      &  0.84            &      &   1.01     &    1.01    \\
 $n_{e}$       &     &   1.30       & 0.78             &      &   1.05     &    1.05 \\
\hline\hline
\end{tabular*}
\end{center}
\end{table}

Figure~\ref{fig:mob} shows the calculated hole-type and electron-type carrier mobilities for both SrPd and BaPd according to Eqs.(1) and (2). At low temperature (10 K), SrPd has a similar hole mobility compared with BaPd, but a much higher electron mobility (2.28 $\times {10^5}$ cm${^2}$V${^{-1}}$s${^{-1}}$). Empirically, the carrier mobility is inversely proportional to the effective mass. We have thus examined the mass of electronic density of states $m_d$ of the valence (VB1) and conduction (CB1) bands in the primitive cell of SrPd (Fig.~\ref{fig:band}), which are 0.203${m_0}$ and 0.177${m_0}$ (Table~\ref{tab:II} with the description in the Appendix), respectively. The tiny effective mass of the conduction band of SrPd may be responsible for its rather high electron mobility. With increasing temperature, the carrier mobilities show dramatical reduction, which is in good accordance with the experimental observations on the temperature-dependent carrier mobilities of XMR materials~\cite{LaSbShanShan2016}.

\begin{figure}[!t]
  \centering
  \includegraphics[width=0.94\columnwidth]{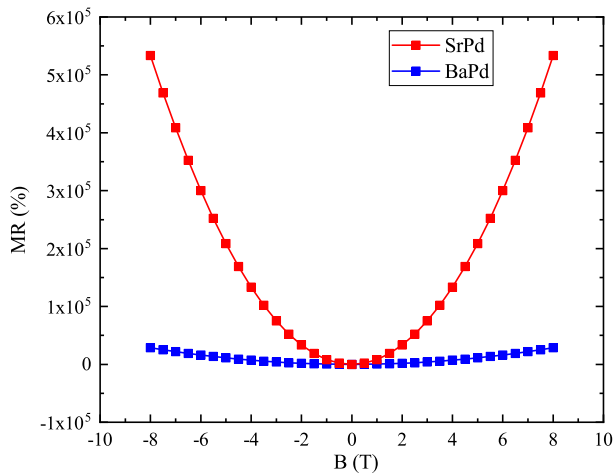}
  \caption{\label{fig:MR}(Color online) Calculated magnetic-field dependence of magnetoresistance for SrPd and BaPd at 10 K.}
\end{figure}

With the calculated carrier mobilities for SrPd and BaPd, we then plot the magnetoresistance as a function of magnetic field in Fig.~\ref{fig:MR} based on Eq. (5). Here, the direction of the magnetic field is perpendicular to the transport direction. As we can see, the magnetoresistance of SrPd under 4 T and 10 K can reach as high as ${10^5}\%$, whereas that of BaPd is in order of 10$^4\%$. Our calculations thus suggest that both SrPd and BaPd would exhibit the XMR effect.

The topological property of SrPd and BaPd has also been examined using the Wilson loops method~\cite{PhysRevB.84.075119, PhysRevB.83.235401}. The calculated topological indexes of both SrPd and BaPd are $(0; 000)$, which suggests that they are both topologically trivial~\cite{PhysRevLett.98.106803}. These findings are consistent with the catalogue of topological materials in previous theoretical studies~\cite{2019Cata-topMater1, 2019Cata-topMater2, 2019Cata-topMater3}.

\section{discussion and Summary}

Both the topological protection and the charge compensation mechanisms have been proposed for the XMR effect. To further ascertain the electron-hole compensation mechanism underlying the XMR effect, we need to find more XMR materials with the following properties: (1) good charge compensation; (2) high carrier mobilities; (3) trivial topological property. Here, based on first-principles electronic structure calculations, we find that SrPd meets the requirements just mentioned. Our calculations show that both the electron-type and hole-type carrier densities of SrPd are in order of {10$^{20}$} cm$^{-3}$, which are in good electron-hole compensation. Meanwhile, the calculated electron-type and hole-type carrier mobilities at low temperature are as high as 2.28$\times$10$^5$ cm$^2$V$^{-1}$s$^{-1}$ and 5.37$\times$10$^4$ cm$^2$V$^{-1}$s$^{-1}$, respectively. Moreover, the calculated topological invariant indicates that SrPd is topologically trivial. Notably, SrPd is a two-band semimetal with only two doubly degenerate bands crossing the Fermi level, which can be more precisely described by the semiclassical two-band model than previous topologically trivial XMR materials LaSb and YSb~\cite{PhysRevB.93.235142, LaSbShanShan2016, PhysRevLett.117.127204, YSbQiaoHe2017}.
According to the two-band model, the good charge compensation with rather high carrier mobilities leads to an XMR of ${10^5}\%$ in SrPd. Likewise, we find that BaPd is also a topologically trivial material with charge compensation, while its relatively lower carrier mobilities than those of SrPd may lead to a smaller MR of ${10^4}\%$.

In summary, our first-principles electronic structure calculations show that both SrPd and BaPd are semimetals with good electron-hole compensation as well as high carrier mobilities. Moreover, there are only two bands crossing the Fermi level. Based on the two-band semiclassical model, we predict that  both SrPd and BaPd will demonstrate extremely large magnetoresistance. The calculated topological indexes also indicate that the electronic band structures of both SrPd and BaPd are topologically trivial. These results suggest that both SrPd and BaPd can serve as an ideal platform to to examine the charge compensation mechanism for the XMR effect with excluding nontrivial topological property, which waits for further experimental verification.

\begin{acknowledgments}
We wish to thank W. Ji, C. Wang, X. H. Kong, H. Y. Lv, H.-C. Yang, and B.-C. Gong for helpful discussions. This work was supported by the National Natural Science Foundation of China (Grants No. 11774424 and No. 11774422), the National Key R$\&$D Program of China (Grant No. 2017YFA0302903), the CAS Interdisciplinary Innovation Team, the Fundamental Research Funds for the Central Universities, and the Research Funds of Renmin University of China (Grants No. 16XNLQ01 and No. 19XNLG13). Computational resources were provided by the Physical Laboratory of High Performance Computing at Renmin University of China.
\end{acknowledgments}

\section{Appendix}
\begin{figure}[!t]
  \centering
  \includegraphics[width=0.98\columnwidth]{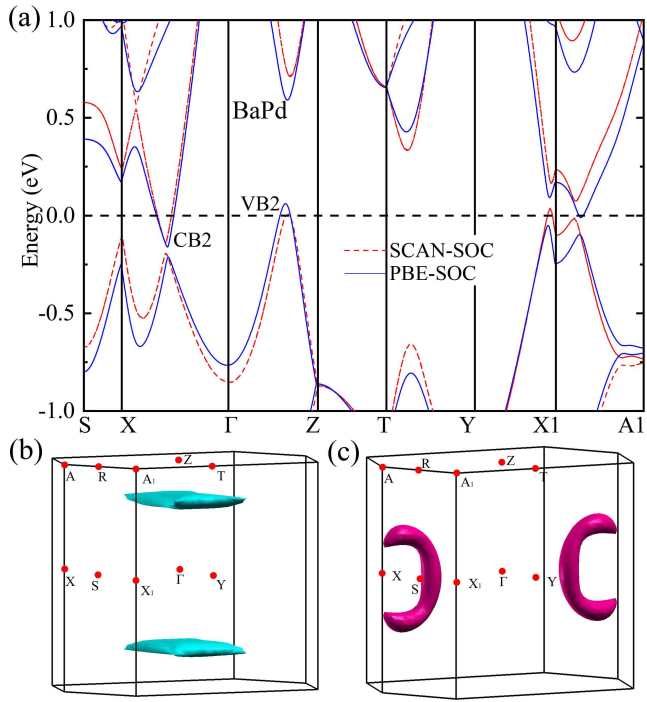}
  \caption{\label{fig:band2}(Color online) (a) Band structures of BaPd along high-symmetry paths of primitive cell calculated with the PBE functional (blue solid line) and the SCAN functional (red dashed line) including the SOC effect. The valence (VB2) and the conduction (CB2) bands are marked. (b) Hole-type and (c) electron-type Fermi pockets of BaPd in primitive cell calculated with the PBE functional.}
\end{figure}

\renewcommand\arraystretch{1.8}
\begin{table}[!b]
\caption{\label{tab:II} Calculated effective mass $m_\beta^*/{m_0}$ along $\beta$ direction and the corresponding mass of electronic density of states $m_d$ for SrPd in the primitive cell. Here $m_0$ is the mass of free electron. }
\begin{center}
\begin{tabular*}{1\columnwidth}{@{\extracolsep{\fill}}ccccc}
\hline\hline
  SrPd        &  $a$  &   $b$  &   $c$   &    $m_d$    \\
\hline
  VB1         &    -0.075    &   -0.039      &  2.844        &     0.203      \\
  CB1         &    0.056      &   0.052       &  1.912         &     0.177  \\
\hline\hline
\end{tabular*}
\end{center}
\end{table}

The band structures along the high-symmetry paths in the BZ of primitive cell and the Fermi surface of BaPd are shown in Fig.~\ref{fig:band2}. Both the results calculated respectively  with the PBE functional at GGA level (blue solid line) and the SCAN functional at meta-GGA level (red dotted line) are presented.

The effective mass is an important indicator for transport properties. If the Fermi surface is anisotropic, the effective mass ${m^*}$ should be replaced by $\sqrt{|{m_\beta ^*m_d^{}}|}$, where $m_\beta ^*={\hbar ^2}/\left[{{\partial ^2}\varepsilon\left({{k_\beta}}\right)/\partial{k_\beta}^2}\right]$ is the effective mass along the transport direction $\beta$ and $m_d^{}=\sqrt[3]{{\left| {m_a^*m_b^*m_c^*}\right|}}$  is the mass of electronic density of states~\cite{PhysRev.80.72, C2NR30585B, Fiori.2013, qiao.nc2014, Zhang2014}.
Generally, the effective mass $m_\beta ^*$ is calculated at the valence band maximum (VBM) for the holes and at the conduction band minimum (CBM) for the electrons. Nevertheless, the global VBM of SrPd is not on the high-symmetry line, but locates at an ordinary point in the reciprocal space \textcolor{blue}{~\cite{PhysRevB.74.205113, PhysRevB.76.075201, Nanotechnology-29-075701}} with the coordinate (0.17, -0.17, 0.12). The highest-energy valence band along the high-symmetry line of primitive cell is marked by VB1 in Fig.~\ref{fig:band}(a). Correspondingly, the CBM locating at (0.70, 0.30, 0.00) along the $Y$-$X1$ line is marked by CB1 in Fig.~\ref{fig:band}(a). Then we select the VB1 for the holes and the CB1 for the electrons to calculate $m_\beta ^*$. The calculated ${{m_\beta ^*}/{m_0}}$ along different directions of primitive cell are presented in Table~\ref{tab:II}.
The ${{m_\beta ^*}/{m_0}}$  for both holes and electrons are in the order of ${10^{ - 2}}$ along the $a$-axis and $b$-axis, but with much larger values along the $c$-axis. The anisotropic effective mass also reflects the anisotropic transport property of SrPd.

\normalem

\begin{thebibliography}{66}%
\makeatletter
\providecommand \@ifxundefined [1]{%
 \@ifx{#1\undefined}
}%
\providecommand \@ifnum [1]{%
 \ifnum #1\expandafter \@firstoftwo
 \else \expandafter \@secondoftwo
 \fi
}%
\providecommand \@ifx [1]{%
 \ifx #1\expandafter \@firstoftwo
 \else \expandafter \@secondoftwo
 \fi
}%
\providecommand \natexlab [1]{#1}%
\providecommand \enquote  [1]{``#1''}%
\providecommand \bibnamefont  [1]{#1}%
\providecommand \bibfnamefont [1]{#1}%
\providecommand \citenamefont [1]{#1}%
\providecommand \href@noop [0]{\@secondoftwo}%
\providecommand \href [0]{\begingroup \@sanitize@url \@href}%
\providecommand \@href[1]{\@@startlink{#1}\@@href}%
\providecommand \@@href[1]{\endgroup#1\@@endlink}%
\providecommand \@sanitize@url [0]{\catcode `\\12\catcode `\$12\catcode
  `\&12\catcode `\#12\catcode `\^12\catcode `\_12\catcode `\%12\relax}%
\providecommand \@@startlink[1]{}%
\providecommand \@@endlink[0]{}%
\providecommand \url  [0]{\begingroup\@sanitize@url \@url }%
\providecommand \@url [1]{\endgroup\@href {#1}{\urlprefix }}%
\providecommand \urlprefix  [0]{URL }%
\providecommand \Eprint [0]{\href }%
\providecommand \doibase [0]{http://dx.doi.org/}%
\providecommand \selectlanguage [0]{\@gobble}%
\providecommand \bibinfo  [0]{\@secondoftwo}%
\providecommand \bibfield  [0]{\@secondoftwo}%
\providecommand \translation [1]{[#1]}%
\providecommand \BibitemOpen [0]{}%
\providecommand \bibitemStop [0]{}%
\providecommand \bibitemNoStop [0]{.\EOS\space}%
\providecommand \EOS [0]{\spacefactor3000\relax}%
\providecommand \BibitemShut  [1]{\csname bibitem#1\endcsname}%
\let\auto@bib@innerbib\@empty
\bibitem [{\citenamefont {Lenz}(1990)}]{lenz1990review}%
  \BibitemOpen
  \bibfield  {author} {\bibinfo {author} {\bibfnamefont {J.~E.}\ \bibnamefont
  {Lenz}},\ }\href {\doibase 10.1109/5.56910} {\bibfield  {journal} {\bibinfo
  {journal} {Proc. IEEE}\ }\textbf {\bibinfo {volume} {78}},\ \bibinfo {pages}
  {973} (\bibinfo {year} {1990})}\BibitemShut {NoStop}%
\bibitem [{\citenamefont {Daughton}(1999)}]{daughton1999gmr}%
  \BibitemOpen
  \bibfield  {author} {\bibinfo {author} {\bibfnamefont {J.}~\bibnamefont
  {Daughton}},\ }\href
  {http://www.sciencedirect.com/science/article/pii/S030488539800376X}
  {\bibfield  {journal} {\bibinfo  {journal} {J. Magn. Magn. Mater.}\ }\textbf
  {\bibinfo {volume} {192}},\ \bibinfo {pages} {334 } (\bibinfo {year}
  {1999})}\BibitemShut {NoStop}%
\bibitem [{\citenamefont {Reig}\ \emph {et~al.}(2009)\citenamefont {Reig},
  \citenamefont {Cubells-Beltr{\'a}n},\ and\ \citenamefont
  {Ram{\'\i}rez~Mu{\~n}oz}}]{reig2009magnetic}%
  \BibitemOpen
  \bibfield  {author} {\bibinfo {author} {\bibfnamefont {C.}~\bibnamefont
  {Reig}}, \bibinfo {author} {\bibfnamefont {M.-D.}\ \bibnamefont
  {Cubells-Beltr{\'a}n}}, \ and\ \bibinfo {author} {\bibfnamefont
  {D.}~\bibnamefont {Ram{\'\i}rez~Mu{\~n}oz}},\ }\href {\doibase
  10.3390/s91007919} {\bibfield  {journal} {\bibinfo  {journal} {Sensors}\
  }\textbf {\bibinfo {volume} {9}},\ \bibinfo {pages} {7919} (\bibinfo {year}
  {2009})}\BibitemShut {NoStop}%
\bibitem [{\citenamefont {McGuire}\ and\ \citenamefont
  {Potter}(1975)}]{mcguire1975anisotropic}%
  \BibitemOpen
  \bibfield  {author} {\bibinfo {author} {\bibfnamefont {T.}~\bibnamefont
  {McGuire}}\ and\ \bibinfo {author} {\bibfnamefont {R.}~\bibnamefont
  {Potter}},\ }\href
  {http://www.unife.it/scienze/lm.fisica/insegnamenti/proprieta-magnetiche-materia/materiale/magnetoresistenza_anisotropa.pdf}
  {\bibfield  {journal} {\bibinfo  {journal} {IEEE Trans. Magn.}\ }\textbf
  {\bibinfo {volume} {11}},\ \bibinfo {pages} {1018} (\bibinfo {year}
  {1975})}\BibitemShut {NoStop}%
\bibitem [{\citenamefont {Baibich}\ \emph {et~al.}(1988)\citenamefont
  {Baibich}, \citenamefont {Broto}, \citenamefont {Fert}, \citenamefont
  {Van~Dau}, \citenamefont {Petroff}, \citenamefont {Etienne}, \citenamefont
  {Creuzet}, \citenamefont {Friederich},\ and\ \citenamefont
  {Chazelas}}]{PhysRevLett.61.2472}%
  \BibitemOpen
  \bibfield  {author} {\bibinfo {author} {\bibfnamefont {M.~N.}\ \bibnamefont
  {Baibich}}, \bibinfo {author} {\bibfnamefont {J.~M.}\ \bibnamefont {Broto}},
  \bibinfo {author} {\bibfnamefont {A.}~\bibnamefont {Fert}}, \bibinfo {author}
  {\bibfnamefont {F.~N.}\ \bibnamefont {Van~Dau}}, \bibinfo {author}
  {\bibfnamefont {F.}~\bibnamefont {Petroff}}, \bibinfo {author} {\bibfnamefont
  {P.}~\bibnamefont {Etienne}}, \bibinfo {author} {\bibfnamefont
  {G.}~\bibnamefont {Creuzet}}, \bibinfo {author} {\bibfnamefont
  {A.}~\bibnamefont {Friederich}}, \ and\ \bibinfo {author} {\bibfnamefont
  {J.}~\bibnamefont {Chazelas}},\ }\href {\doibase 10.1103/PhysRevLett.61.2472}
  {\bibfield  {journal} {\bibinfo  {journal} {Phys. Rev. Lett.}\ }\textbf
  {\bibinfo {volume} {61}},\ \bibinfo {pages} {2472} (\bibinfo {year}
  {1988})}\BibitemShut {NoStop}%
\bibitem [{\citenamefont {Dagotto}\ \emph {et~al.}(2001)\citenamefont
  {Dagotto}, \citenamefont {Hotta},\ and\ \citenamefont
  {Moreo}}]{dagotto2001colossal}%
  \BibitemOpen
  \bibfield  {author} {\bibinfo {author} {\bibfnamefont {E.}~\bibnamefont
  {Dagotto}}, \bibinfo {author} {\bibfnamefont {T.}~\bibnamefont {Hotta}}, \
  and\ \bibinfo {author} {\bibfnamefont {A.}~\bibnamefont {Moreo}},\ }\href
  {\doibase https://doi.org/10.1016/S0370-1573(00)00121-6} {\bibfield
  {journal} {\bibinfo  {journal} {Phys. Rep.}\ }\textbf {\bibinfo {volume}
  {344}},\ \bibinfo {pages} {1 } (\bibinfo {year} {2001})}\BibitemShut
  {NoStop}%
\bibitem [{\citenamefont {Salamon}\ and\ \citenamefont
  {Jaime}(2001)}]{salamon2001physics}%
  \BibitemOpen
  \bibfield  {author} {\bibinfo {author} {\bibfnamefont {M.~B.}\ \bibnamefont
  {Salamon}}\ and\ \bibinfo {author} {\bibfnamefont {M.}~\bibnamefont
  {Jaime}},\ }\href {\doibase 10.1103/RevModPhys.73.583} {\bibfield  {journal}
  {\bibinfo  {journal} {Rev. Mod. Phys.}\ }\textbf {\bibinfo {volume} {73}},\
  \bibinfo {pages} {583} (\bibinfo {year} {2001})}\BibitemShut {NoStop}%
\bibitem [{\citenamefont {Moodera}\ \emph {et~al.}(1995)\citenamefont
  {Moodera}, \citenamefont {Kinder}, \citenamefont {Wong},\ and\ \citenamefont
  {Meservey}}]{moodera1995large}%
  \BibitemOpen
  \bibfield  {author} {\bibinfo {author} {\bibfnamefont {J.~S.}\ \bibnamefont
  {Moodera}}, \bibinfo {author} {\bibfnamefont {L.~R.}\ \bibnamefont {Kinder}},
  \bibinfo {author} {\bibfnamefont {T.~M.}\ \bibnamefont {Wong}}, \ and\
  \bibinfo {author} {\bibfnamefont {R.}~\bibnamefont {Meservey}},\ }\href
  {\doibase 10.1103/PhysRevLett.74.3273} {\bibfield  {journal} {\bibinfo
  {journal} {Phys. Rev. Lett.}\ }\textbf {\bibinfo {volume} {74}},\ \bibinfo
  {pages} {3273} (\bibinfo {year} {1995})}\BibitemShut {NoStop}%
\bibitem [{\citenamefont {Parkin}\ \emph {et~al.}(2004)\citenamefont {Parkin},
  \citenamefont {Kaiser}, \citenamefont {Panchula}, \citenamefont {Rice},
  \citenamefont {Hughes}, \citenamefont {Samant},\ and\ \citenamefont
  {Yang}}]{parkin2004giant}%
  \BibitemOpen
  \bibfield  {author} {\bibinfo {author} {\bibfnamefont {S.~S.~P.}\
  \bibnamefont {Parkin}}, \bibinfo {author} {\bibfnamefont {C.}~\bibnamefont
  {Kaiser}}, \bibinfo {author} {\bibfnamefont {A.}~\bibnamefont {Panchula}},
  \bibinfo {author} {\bibfnamefont {P.~M.}\ \bibnamefont {Rice}}, \bibinfo
  {author} {\bibfnamefont {B.}~\bibnamefont {Hughes}}, \bibinfo {author}
  {\bibfnamefont {M.}~\bibnamefont {Samant}}, \ and\ \bibinfo {author}
  {\bibfnamefont {S.-H.}\ \bibnamefont {Yang}},\ }\href {\doibase
  10.1038/nmat1256} {\bibfield  {journal} {\bibinfo  {journal} {Nat. Mater.}\
  }\textbf {\bibinfo {volume} {3}},\ \bibinfo {pages} {862} (\bibinfo {year}
  {2004})}\BibitemShut {NoStop}%
\bibitem [{\citenamefont {Pippard}(1989)}]{pippard1989magnetoresistance}%
  \BibitemOpen
  \bibfield  {author} {\bibinfo {author} {\bibfnamefont {A.~B.}\ \bibnamefont
  {Pippard}},\ }\href@noop {} {\emph {\bibinfo {title} {Magnetoresistance in
  metals}}}\ (\bibinfo  {publisher} {Cambridge University Press, UK},\ \bibinfo
  {year} {1989})\BibitemShut {NoStop}%
\bibitem [{\citenamefont {St{\"o}hr}\ and\ \citenamefont
  {Siegmann}(2006)}]{stohr2006and}%
  \BibitemOpen
  \bibfield  {author} {\bibinfo {author} {\bibfnamefont {J.}~\bibnamefont
  {St{\"o}hr}}\ and\ \bibinfo {author} {\bibfnamefont {H.~C.}\ \bibnamefont
  {Siegmann}},\ }\href@noop {} {\emph {\bibinfo {title} {Magnetism: from
  fundamentals to nanoscale dynamics}}}\ (\bibinfo  {publisher} {Springer
  Science \& Business Media},\ \bibinfo {year} {2006})\BibitemShut {NoStop}%
\bibitem [{\citenamefont {Tafti}\ \emph {et~al.}(2016)\citenamefont {Tafti},
  \citenamefont {Gibson}, \citenamefont {Kushwaha}, \citenamefont
  {Haldolaarachchige},\ and\ \citenamefont {Cava}}]{tafti2016resistivity}%
  \BibitemOpen
  \bibfield  {author} {\bibinfo {author} {\bibfnamefont {F.}~\bibnamefont
  {Tafti}}, \bibinfo {author} {\bibfnamefont {Q.}~\bibnamefont {Gibson}},
  \bibinfo {author} {\bibfnamefont {S.}~\bibnamefont {Kushwaha}}, \bibinfo
  {author} {\bibfnamefont {N.}~\bibnamefont {Haldolaarachchige}}, \ and\
  \bibinfo {author} {\bibfnamefont {R.}~\bibnamefont {Cava}},\ }\href {\doibase
  10.1038/nphys3581} {\bibfield  {journal} {\bibinfo  {journal} {Nat. Phys.}\
  }\textbf {\bibinfo {volume} {12}},\ \bibinfo {pages} {272} (\bibinfo {year}
  {2016})}\BibitemShut {NoStop}%
\bibitem [{\citenamefont {Sun}\ \emph {et~al.}(2016)\citenamefont {Sun},
  \citenamefont {Wang}, \citenamefont {Guo}, \citenamefont {Liu},\ and\
  \citenamefont {Lei}}]{LaSbShanShan2016}%
  \BibitemOpen
  \bibfield  {author} {\bibinfo {author} {\bibfnamefont {S.}~\bibnamefont
  {Sun}}, \bibinfo {author} {\bibfnamefont {Q.}~\bibnamefont {Wang}}, \bibinfo
  {author} {\bibfnamefont {P.-J.}\ \bibnamefont {Guo}}, \bibinfo {author}
  {\bibfnamefont {K.}~\bibnamefont {Liu}}, \ and\ \bibinfo {author}
  {\bibfnamefont {H.}~\bibnamefont {Lei}},\ }\href
  {http://stacks.iop.org/1367-2630/18/i=8/a=082002} {\bibfield  {journal}
  {\bibinfo  {journal} {New J. Phys.}\ }\textbf {\bibinfo {volume} {18}},\
  \bibinfo {pages} {082002} (\bibinfo {year} {2016})}\BibitemShut {NoStop}%
\bibitem [{\citenamefont {Guo}\ \emph {et~al.}(2016)\citenamefont {Guo},
  \citenamefont {Yang}, \citenamefont {Zhang}, \citenamefont {Liu},\ and\
  \citenamefont {Lu}}]{PhysRevB.93.235142}%
  \BibitemOpen
  \bibfield  {author} {\bibinfo {author} {\bibfnamefont {P.-J.}\ \bibnamefont
  {Guo}}, \bibinfo {author} {\bibfnamefont {H.-C.}\ \bibnamefont {Yang}},
  \bibinfo {author} {\bibfnamefont {B.-J.}\ \bibnamefont {Zhang}}, \bibinfo
  {author} {\bibfnamefont {K.}~\bibnamefont {Liu}}, \ and\ \bibinfo {author}
  {\bibfnamefont {Z.-Y.}\ \bibnamefont {Lu}},\ }\href {\doibase
  10.1103/PhysRevB.93.235142} {\bibfield  {journal} {\bibinfo  {journal} {Phys.
  Rev. B}\ }\textbf {\bibinfo {volume} {93}},\ \bibinfo {pages} {235142}
  (\bibinfo {year} {2016})}\BibitemShut {NoStop}%
\bibitem [{\citenamefont {Zeng}\ \emph {et~al.}(2016)\citenamefont {Zeng},
  \citenamefont {Lou}, \citenamefont {Wu}, \citenamefont {Xu}, \citenamefont
  {Guo}, \citenamefont {Kong}, \citenamefont {Zhong}, \citenamefont {Ma},
  \citenamefont {Fu}, \citenamefont {Richard}, \citenamefont {Wang},
  \citenamefont {Liu}, \citenamefont {Lu}, \citenamefont {Huang}, \citenamefont
  {Fang}, \citenamefont {Sun}, \citenamefont {Wang}, \citenamefont {Wang},
  \citenamefont {Shi}, \citenamefont {Weng}, \citenamefont {Lei}, \citenamefont
  {Liu}, \citenamefont {Wang}, \citenamefont {Qian}, \citenamefont {Luo},\ and\
  \citenamefont {Ding}}]{PhysRevLett.117.127204}%
  \BibitemOpen
  \bibfield  {author} {\bibinfo {author} {\bibfnamefont {L.-K.}\ \bibnamefont
  {Zeng}}, \bibinfo {author} {\bibfnamefont {R.}~\bibnamefont {Lou}}, \bibinfo
  {author} {\bibfnamefont {D.-S.}\ \bibnamefont {Wu}}, \bibinfo {author}
  {\bibfnamefont {Q.~N.}\ \bibnamefont {Xu}}, \bibinfo {author} {\bibfnamefont
  {P.-J.}\ \bibnamefont {Guo}}, \bibinfo {author} {\bibfnamefont {L.-Y.}\
  \bibnamefont {Kong}}, \bibinfo {author} {\bibfnamefont {Y.-G.}\ \bibnamefont
  {Zhong}}, \bibinfo {author} {\bibfnamefont {J.-Z.}\ \bibnamefont {Ma}},
  \bibinfo {author} {\bibfnamefont {B.-B.}\ \bibnamefont {Fu}}, \bibinfo
  {author} {\bibfnamefont {P.}~\bibnamefont {Richard}}, \bibinfo {author}
  {\bibfnamefont {P.}~\bibnamefont {Wang}}, \bibinfo {author} {\bibfnamefont
  {G.~T.}\ \bibnamefont {Liu}}, \bibinfo {author} {\bibfnamefont
  {L.}~\bibnamefont {Lu}}, \bibinfo {author} {\bibfnamefont {Y.-B.}\
  \bibnamefont {Huang}}, \bibinfo {author} {\bibfnamefont {C.}~\bibnamefont
  {Fang}}, \bibinfo {author} {\bibfnamefont {S.-S.}\ \bibnamefont {Sun}},
  \bibinfo {author} {\bibfnamefont {Q.}~\bibnamefont {Wang}}, \bibinfo {author}
  {\bibfnamefont {L.}~\bibnamefont {Wang}}, \bibinfo {author} {\bibfnamefont
  {Y.-G.}\ \bibnamefont {Shi}}, \bibinfo {author} {\bibfnamefont {H.~M.}\
  \bibnamefont {Weng}}, \bibinfo {author} {\bibfnamefont {H.-C.}\ \bibnamefont
  {Lei}}, \bibinfo {author} {\bibfnamefont {K.}~\bibnamefont {Liu}}, \bibinfo
  {author} {\bibfnamefont {S.-C.}\ \bibnamefont {Wang}}, \bibinfo {author}
  {\bibfnamefont {T.}~\bibnamefont {Qian}}, \bibinfo {author} {\bibfnamefont
  {J.-L.}\ \bibnamefont {Luo}}, \ and\ \bibinfo {author} {\bibfnamefont
  {H.}~\bibnamefont {Ding}},\ }\href {\doibase 10.1103/PhysRevLett.117.127204}
  {\bibfield  {journal} {\bibinfo  {journal} {Phys. Rev. Lett.}\ }\textbf
  {\bibinfo {volume} {117}},\ \bibinfo {pages} {127204} (\bibinfo {year}
  {2016})}\BibitemShut {NoStop}%
\bibitem [{\citenamefont {Yu}\ \emph {et~al.}(2017)\citenamefont {Yu},
  \citenamefont {Wang}, \citenamefont {Lou}, \citenamefont {Guo}, \citenamefont
  {Xu}, \citenamefont {Liu}, \citenamefont {Wang},\ and\ \citenamefont
  {Xia}}]{YSbQiaoHe2017}%
  \BibitemOpen
  \bibfield  {author} {\bibinfo {author} {\bibfnamefont {Q.-H.}\ \bibnamefont
  {Yu}}, \bibinfo {author} {\bibfnamefont {Y.-Y.}\ \bibnamefont {Wang}},
  \bibinfo {author} {\bibfnamefont {R.}~\bibnamefont {Lou}}, \bibinfo {author}
  {\bibfnamefont {P.-J.}\ \bibnamefont {Guo}}, \bibinfo {author} {\bibfnamefont
  {S.}~\bibnamefont {Xu}}, \bibinfo {author} {\bibfnamefont {K.}~\bibnamefont
  {Liu}}, \bibinfo {author} {\bibfnamefont {S.}~\bibnamefont {Wang}}, \ and\
  \bibinfo {author} {\bibfnamefont {T.-L.}\ \bibnamefont {Xia}},\ }\href
  {http://stacks.iop.org/0295-5075/119/i=1/a=17002} {\bibfield  {journal}
  {\bibinfo  {journal} {Europhys. Lett.}\ }\textbf {\bibinfo {volume} {119}},\
  \bibinfo {pages} {17002} (\bibinfo {year} {2017})}\BibitemShut {NoStop}%
\bibitem [{\citenamefont {He}\ \emph {et~al.}(2016)\citenamefont {He},
  \citenamefont {Zhang}, \citenamefont {Ghimire}, \citenamefont {Liang},
  \citenamefont {Jia}, \citenamefont {Jiang}, \citenamefont {Tang},
  \citenamefont {Chen}, \citenamefont {He}, \citenamefont {Mo}, \citenamefont
  {Hwang}, \citenamefont {Hashimoto}, \citenamefont {Lu}, \citenamefont
  {Moritz}, \citenamefont {Devereaux}, \citenamefont {Chen}, \citenamefont
  {Mitchell},\ and\ \citenamefont {Shen}}]{PhysRevLett.117.267201}%
  \BibitemOpen
  \bibfield  {author} {\bibinfo {author} {\bibfnamefont {J.}~\bibnamefont
  {He}}, \bibinfo {author} {\bibfnamefont {C.}~\bibnamefont {Zhang}}, \bibinfo
  {author} {\bibfnamefont {N.~J.}\ \bibnamefont {Ghimire}}, \bibinfo {author}
  {\bibfnamefont {T.}~\bibnamefont {Liang}}, \bibinfo {author} {\bibfnamefont
  {C.}~\bibnamefont {Jia}}, \bibinfo {author} {\bibfnamefont {J.}~\bibnamefont
  {Jiang}}, \bibinfo {author} {\bibfnamefont {S.}~\bibnamefont {Tang}},
  \bibinfo {author} {\bibfnamefont {S.}~\bibnamefont {Chen}}, \bibinfo {author}
  {\bibfnamefont {Y.}~\bibnamefont {He}}, \bibinfo {author} {\bibfnamefont
  {S.-K.}\ \bibnamefont {Mo}}, \bibinfo {author} {\bibfnamefont {C.~C.}\
  \bibnamefont {Hwang}}, \bibinfo {author} {\bibfnamefont {M.}~\bibnamefont
  {Hashimoto}}, \bibinfo {author} {\bibfnamefont {D.~H.}\ \bibnamefont {Lu}},
  \bibinfo {author} {\bibfnamefont {B.}~\bibnamefont {Moritz}}, \bibinfo
  {author} {\bibfnamefont {T.~P.}\ \bibnamefont {Devereaux}}, \bibinfo {author}
  {\bibfnamefont {Y.~L.}\ \bibnamefont {Chen}}, \bibinfo {author}
  {\bibfnamefont {J.~F.}\ \bibnamefont {Mitchell}}, \ and\ \bibinfo {author}
  {\bibfnamefont {Z.-X.}\ \bibnamefont {Shen}},\ }\href {\doibase
  10.1103/PhysRevLett.117.267201} {\bibfield  {journal} {\bibinfo  {journal}
  {Phys. Rev. Lett.}\ }\textbf {\bibinfo {volume} {117}},\ \bibinfo {pages}
  {267201} (\bibinfo {year} {2016})}\BibitemShut {NoStop}%
\bibitem [{\citenamefont {Liang}\ \emph {et~al.}(2015)\citenamefont {Liang},
  \citenamefont {Gibson}, \citenamefont {Ali}, \citenamefont {Liu},
  \citenamefont {Cava},\ and\ \citenamefont {Ong}}]{liang2015ultrahigh}%
  \BibitemOpen
  \bibfield  {author} {\bibinfo {author} {\bibfnamefont {T.}~\bibnamefont
  {Liang}}, \bibinfo {author} {\bibfnamefont {Q.}~\bibnamefont {Gibson}},
  \bibinfo {author} {\bibfnamefont {M.~N.}\ \bibnamefont {Ali}}, \bibinfo
  {author} {\bibfnamefont {M.}~\bibnamefont {Liu}}, \bibinfo {author}
  {\bibfnamefont {R.}~\bibnamefont {Cava}}, \ and\ \bibinfo {author}
  {\bibfnamefont {N.}~\bibnamefont {Ong}},\ }\href
  {http://dx.doi.org/10.1038/nmat4143} {\bibfield  {journal} {\bibinfo
  {journal} {Nat. Mater.}\ }\textbf {\bibinfo {volume} {14}},\ \bibinfo {pages}
  {280} (\bibinfo {year} {2015})}\BibitemShut {NoStop}%
\bibitem [{\citenamefont {Wang}\ \emph {et~al.}(2014)\citenamefont {Wang},
  \citenamefont {Graf}, \citenamefont {Li}, \citenamefont {Wang},\ and\
  \citenamefont {Petrovic}}]{wang2014anisotropic}%
  \BibitemOpen
  \bibfield  {author} {\bibinfo {author} {\bibfnamefont {K.}~\bibnamefont
  {Wang}}, \bibinfo {author} {\bibfnamefont {D.}~\bibnamefont {Graf}}, \bibinfo
  {author} {\bibfnamefont {L.}~\bibnamefont {Li}}, \bibinfo {author}
  {\bibfnamefont {L.}~\bibnamefont {Wang}}, \ and\ \bibinfo {author}
  {\bibfnamefont {C.}~\bibnamefont {Petrovic}},\ }\href
  {http://dx.doi.org/10.1038/srep07328} {\bibfield  {journal} {\bibinfo
  {journal} {Sci. Rep.}\ }\textbf {\bibinfo {volume} {4}},\ \bibinfo {pages}
  {7328} (\bibinfo {year} {2014})}\BibitemShut {NoStop}%
\bibitem [{\citenamefont {Ali}\ \emph {et~al.}(2014)\citenamefont {Ali},
  \citenamefont {Xiong}, \citenamefont {Flynn}, \citenamefont {Tao},
  \citenamefont {Gibson}, \citenamefont {Schoop}, \citenamefont {Liang},
  \citenamefont {Haldolaarachchige}, \citenamefont {Hirschberger},
  \citenamefont {Ong},\ and\ \citenamefont {Cava}}]{ali2014large}%
  \BibitemOpen
  \bibfield  {author} {\bibinfo {author} {\bibfnamefont {M.~N.}\ \bibnamefont
  {Ali}}, \bibinfo {author} {\bibfnamefont {J.}~\bibnamefont {Xiong}}, \bibinfo
  {author} {\bibfnamefont {S.}~\bibnamefont {Flynn}}, \bibinfo {author}
  {\bibfnamefont {J.}~\bibnamefont {Tao}}, \bibinfo {author} {\bibfnamefont
  {Q.~D.}\ \bibnamefont {Gibson}}, \bibinfo {author} {\bibfnamefont {L.~M.}\
  \bibnamefont {Schoop}}, \bibinfo {author} {\bibfnamefont {T.}~\bibnamefont
  {Liang}}, \bibinfo {author} {\bibfnamefont {N.}~\bibnamefont
  {Haldolaarachchige}}, \bibinfo {author} {\bibfnamefont {M.}~\bibnamefont
  {Hirschberger}}, \bibinfo {author} {\bibfnamefont {N.}~\bibnamefont {Ong}}, \
  and\ \bibinfo {author} {\bibfnamefont {R.~J.}\ \bibnamefont {Cava}},\ }\href
  {\doibase 10.1038/nature13763} {\bibfield  {journal} {\bibinfo  {journal}
  {Nature}\ }\textbf {\bibinfo {volume} {514}},\ \bibinfo {pages} {205}
  (\bibinfo {year} {2014})}\BibitemShut {NoStop}%
\bibitem [{\citenamefont {Lv}\ \emph {et~al.}(2015)\citenamefont {Lv},
  \citenamefont {Lu}, \citenamefont {Shao}, \citenamefont {Liu}, \citenamefont
  {Tan},\ and\ \citenamefont {Sun}}]{LvWTe2}%
  \BibitemOpen
  \bibfield  {author} {\bibinfo {author} {\bibfnamefont {H.~Y.}\ \bibnamefont
  {Lv}}, \bibinfo {author} {\bibfnamefont {W.~J.}\ \bibnamefont {Lu}}, \bibinfo
  {author} {\bibfnamefont {D.~F.}\ \bibnamefont {Shao}}, \bibinfo {author}
  {\bibfnamefont {Y.}~\bibnamefont {Liu}}, \bibinfo {author} {\bibfnamefont
  {S.~G.}\ \bibnamefont {Tan}}, \ and\ \bibinfo {author} {\bibfnamefont
  {Y.~P.}\ \bibnamefont {Sun}},\ }\href
  {http://stacks.iop.org/0295-5075/110/i=3/a=37004} {\bibfield  {journal}
  {\bibinfo  {journal} {Europhys. Lett.}\ }\textbf {\bibinfo {volume} {110}},\
  \bibinfo {pages} {37004} (\bibinfo {year} {2015})}\BibitemShut {NoStop}%
\bibitem [{\citenamefont {Jiang}\ \emph {et~al.}(2015)\citenamefont {Jiang},
  \citenamefont {Tang}, \citenamefont {Pan}, \citenamefont {Liu}, \citenamefont
  {Niu}, \citenamefont {Wang}, \citenamefont {Xu}, \citenamefont {Yang},
  \citenamefont {Xie}, \citenamefont {Song}, \citenamefont {Dudin},
  \citenamefont {Kim}, \citenamefont {Hoesch}, \citenamefont {Das},
  \citenamefont {Vobornik}, \citenamefont {Wan},\ and\ \citenamefont
  {Feng}}]{2015PRL.115.WTe2SOC}%
  \BibitemOpen
  \bibfield  {author} {\bibinfo {author} {\bibfnamefont {J.}~\bibnamefont
  {Jiang}}, \bibinfo {author} {\bibfnamefont {F.}~\bibnamefont {Tang}},
  \bibinfo {author} {\bibfnamefont {X.~C.}\ \bibnamefont {Pan}}, \bibinfo
  {author} {\bibfnamefont {H.~M.}\ \bibnamefont {Liu}}, \bibinfo {author}
  {\bibfnamefont {X.~H.}\ \bibnamefont {Niu}}, \bibinfo {author} {\bibfnamefont
  {Y.~X.}\ \bibnamefont {Wang}}, \bibinfo {author} {\bibfnamefont {D.~F.}\
  \bibnamefont {Xu}}, \bibinfo {author} {\bibfnamefont {H.~F.}\ \bibnamefont
  {Yang}}, \bibinfo {author} {\bibfnamefont {B.~P.}\ \bibnamefont {Xie}},
  \bibinfo {author} {\bibfnamefont {F.~Q.}\ \bibnamefont {Song}}, \bibinfo
  {author} {\bibfnamefont {P.}~\bibnamefont {Dudin}}, \bibinfo {author}
  {\bibfnamefont {T.~K.}\ \bibnamefont {Kim}}, \bibinfo {author} {\bibfnamefont
  {M.}~\bibnamefont {Hoesch}}, \bibinfo {author} {\bibfnamefont {P.~K.}\
  \bibnamefont {Das}}, \bibinfo {author} {\bibfnamefont {I.}~\bibnamefont
  {Vobornik}}, \bibinfo {author} {\bibfnamefont {X.~G.}\ \bibnamefont {Wan}}, \
  and\ \bibinfo {author} {\bibfnamefont {D.~L.}\ \bibnamefont {Feng}},\ }\href
  {\doibase 10.1103/PhysRevLett.115.166601} {\bibfield  {journal} {\bibinfo
  {journal} {Phys. Rev. Lett.}\ }\textbf {\bibinfo {volume} {115}},\ \bibinfo
  {pages} {166601} (\bibinfo {year} {2015})}\BibitemShut {NoStop}%
\bibitem [{\citenamefont {Huang}\ \emph {et~al.}(2015)\citenamefont {Huang},
  \citenamefont {Zhao}, \citenamefont {Long}, \citenamefont {Wang},
  \citenamefont {Chen}, \citenamefont {Yang}, \citenamefont {Liang},
  \citenamefont {Xue}, \citenamefont {Weng}, \citenamefont {Fang},
  \citenamefont {Dai},\ and\ \citenamefont {Chen}}]{PhysRevX.5.031023}%
  \BibitemOpen
  \bibfield  {author} {\bibinfo {author} {\bibfnamefont {X.}~\bibnamefont
  {Huang}}, \bibinfo {author} {\bibfnamefont {L.}~\bibnamefont {Zhao}},
  \bibinfo {author} {\bibfnamefont {Y.}~\bibnamefont {Long}}, \bibinfo {author}
  {\bibfnamefont {P.}~\bibnamefont {Wang}}, \bibinfo {author} {\bibfnamefont
  {D.}~\bibnamefont {Chen}}, \bibinfo {author} {\bibfnamefont {Z.}~\bibnamefont
  {Yang}}, \bibinfo {author} {\bibfnamefont {H.}~\bibnamefont {Liang}},
  \bibinfo {author} {\bibfnamefont {M.}~\bibnamefont {Xue}}, \bibinfo {author}
  {\bibfnamefont {H.}~\bibnamefont {Weng}}, \bibinfo {author} {\bibfnamefont
  {Z.}~\bibnamefont {Fang}}, \bibinfo {author} {\bibfnamefont {X.}~\bibnamefont
  {Dai}}, \ and\ \bibinfo {author} {\bibfnamefont {G.}~\bibnamefont {Chen}},\
  }\href {\doibase 10.1103/PhysRevX.5.031023} {\bibfield  {journal} {\bibinfo
  {journal} {Phys. Rev. X}\ }\textbf {\bibinfo {volume} {5}},\ \bibinfo {pages}
  {031023} (\bibinfo {year} {2015})}\BibitemShut {NoStop}%
\bibitem [{\citenamefont {Zhang}\ \emph {et~al.}(2018)\citenamefont {Zhang},
  \citenamefont {Wu}, \citenamefont {Liu},\ and\ \citenamefont
  {Yazyev}}]{ZhangSN-arxiv}%
  \BibitemOpen
  \bibfield  {author} {\bibinfo {author} {\bibfnamefont {S.~N.}\ \bibnamefont
  {Zhang}}, \bibinfo {author} {\bibfnamefont {Q.~S.}\ \bibnamefont {Wu}},
  \bibinfo {author} {\bibfnamefont {Y.}~\bibnamefont {Liu}}, \ and\ \bibinfo
  {author} {\bibfnamefont {O.~V.}\ \bibnamefont {Yazyev}},\ }\href
  {https://arxiv.org/abs/1808.08178} {\bibfield  {journal} {\bibinfo  {journal}
  {arXiv:1808.08178}\ } (\bibinfo {year} {2018})}\BibitemShut {NoStop}%
\bibitem [{\citenamefont {Shekhar}\ \emph {et~al.}(2015)\citenamefont
  {Shekhar}, \citenamefont {Nayak}, \citenamefont {Sun}, \citenamefont
  {Schmidt}, \citenamefont {Nicklas}, \citenamefont {Leermakers}, \citenamefont
  {Zeitler}, \citenamefont {Skourski}, \citenamefont {Wosnitza}, \citenamefont
  {Liu} \emph {et~al.}}]{shekhar2015extremely}%
  \BibitemOpen
  \bibfield  {author} {\bibinfo {author} {\bibfnamefont {C.}~\bibnamefont
  {Shekhar}}, \bibinfo {author} {\bibfnamefont {A.~K.}\ \bibnamefont {Nayak}},
  \bibinfo {author} {\bibfnamefont {Y.}~\bibnamefont {Sun}}, \bibinfo {author}
  {\bibfnamefont {M.}~\bibnamefont {Schmidt}}, \bibinfo {author} {\bibfnamefont
  {M.}~\bibnamefont {Nicklas}}, \bibinfo {author} {\bibfnamefont
  {I.}~\bibnamefont {Leermakers}}, \bibinfo {author} {\bibfnamefont
  {U.}~\bibnamefont {Zeitler}}, \bibinfo {author} {\bibfnamefont
  {Y.}~\bibnamefont {Skourski}}, \bibinfo {author} {\bibfnamefont
  {J.}~\bibnamefont {Wosnitza}}, \bibinfo {author} {\bibfnamefont
  {Z.}~\bibnamefont {Liu}},  \emph {et~al.},\ }\href {\doibase
  10.1038/nphys3372} {\bibfield  {journal} {\bibinfo  {journal} {Nat. Phys.}\
  }\textbf {\bibinfo {volume} {11}},\ \bibinfo {pages} {645} (\bibinfo {year}
  {2015})}\BibitemShut {NoStop}%
\bibitem [{\citenamefont {Luo}\ \emph {et~al.}(2015)\citenamefont {Luo},
  \citenamefont {Ghimire}, \citenamefont {Wartenbe}, \citenamefont {Choi},
  \citenamefont {Neupane}, \citenamefont {McDonald}, \citenamefont {Bauer},
  \citenamefont {Zhu}, \citenamefont {Thompson},\ and\ \citenamefont
  {Ronning}}]{PhysRevB.92.205134}%
  \BibitemOpen
  \bibfield  {author} {\bibinfo {author} {\bibfnamefont {Y.}~\bibnamefont
  {Luo}}, \bibinfo {author} {\bibfnamefont {N.~J.}\ \bibnamefont {Ghimire}},
  \bibinfo {author} {\bibfnamefont {M.}~\bibnamefont {Wartenbe}}, \bibinfo
  {author} {\bibfnamefont {H.}~\bibnamefont {Choi}}, \bibinfo {author}
  {\bibfnamefont {M.}~\bibnamefont {Neupane}}, \bibinfo {author} {\bibfnamefont
  {R.~D.}\ \bibnamefont {McDonald}}, \bibinfo {author} {\bibfnamefont {E.~D.}\
  \bibnamefont {Bauer}}, \bibinfo {author} {\bibfnamefont {J.}~\bibnamefont
  {Zhu}}, \bibinfo {author} {\bibfnamefont {J.~D.}\ \bibnamefont {Thompson}}, \
  and\ \bibinfo {author} {\bibfnamefont {F.}~\bibnamefont {Ronning}},\ }\href
  {\doibase 10.1103/PhysRevB.92.205134} {\bibfield  {journal} {\bibinfo
  {journal} {Phys. Rev. B}\ }\textbf {\bibinfo {volume} {92}},\ \bibinfo
  {pages} {205134} (\bibinfo {year} {2015})}\BibitemShut {NoStop}%
\bibitem [{\citenamefont {Pletikosi\ifmmode~\acute{c}\else \'{c}\fi{}}\ \emph
  {et~al.}(2014)\citenamefont {Pletikosi\ifmmode~\acute{c}\else \'{c}\fi{}},
  \citenamefont {Ali}, \citenamefont {Fedorov}, \citenamefont {Cava},\ and\
  \citenamefont {Valla}}]{PhysRevLett.113.216601}%
  \BibitemOpen
  \bibfield  {author} {\bibinfo {author} {\bibfnamefont {I.}~\bibnamefont
  {Pletikosi\ifmmode~\acute{c}\else \'{c}\fi{}}}, \bibinfo {author}
  {\bibfnamefont {M.~N.}\ \bibnamefont {Ali}}, \bibinfo {author} {\bibfnamefont
  {A.~V.}\ \bibnamefont {Fedorov}}, \bibinfo {author} {\bibfnamefont {R.~J.}\
  \bibnamefont {Cava}}, \ and\ \bibinfo {author} {\bibfnamefont
  {T.}~\bibnamefont {Valla}},\ }\href {\doibase 10.1103/PhysRevLett.113.216601}
  {\bibfield  {journal} {\bibinfo  {journal} {Phys. Rev. Lett.}\ }\textbf
  {\bibinfo {volume} {113}},\ \bibinfo {pages} {216601} (\bibinfo {year}
  {2014})}\BibitemShut {NoStop}%
\bibitem [{\citenamefont {Kumar}\ \emph {et~al.}(2017)\citenamefont {Kumar},
  \citenamefont {Sun}, \citenamefont {Xu}, \citenamefont {Manna}, \citenamefont
  {Yao}, \citenamefont {S\"uss}, \citenamefont {Leermakers}, \citenamefont
  {Young}, \citenamefont {F\"orster}, \citenamefont {Schmidt}, \citenamefont
  {Zeitler}, \citenamefont {Shi}, \citenamefont {Felser},\ and\ \citenamefont
  {Shekhar}}]{2017WP2-ncomn}%
  \BibitemOpen
  \bibfield  {author} {\bibinfo {author} {\bibfnamefont {N.}~\bibnamefont
  {Kumar}}, \bibinfo {author} {\bibfnamefont {Y.}~\bibnamefont {Sun}}, \bibinfo
  {author} {\bibfnamefont {N.}~\bibnamefont {Xu}}, \bibinfo {author}
  {\bibfnamefont {K.}~\bibnamefont {Manna}}, \bibinfo {author} {\bibfnamefont
  {M.}~\bibnamefont {Yao}}, \bibinfo {author} {\bibfnamefont {V.}~\bibnamefont
  {S\"uss}}, \bibinfo {author} {\bibfnamefont {I.}~\bibnamefont {Leermakers}},
  \bibinfo {author} {\bibfnamefont {O.}~\bibnamefont {Young}}, \bibinfo
  {author} {\bibfnamefont {T.}~\bibnamefont {F\"orster}}, \bibinfo {author}
  {\bibfnamefont {H.}~\bibnamefont {Schmidt}, \bibfnamefont {Marcus~Borrmann}},
  \bibinfo {author} {\bibfnamefont {U.}~\bibnamefont {Zeitler}}, \bibinfo
  {author} {\bibfnamefont {M.}~\bibnamefont {Shi}}, \bibinfo {author}
  {\bibfnamefont {C.}~\bibnamefont {Felser}}, \ and\ \bibinfo {author}
  {\bibfnamefont {C.}~\bibnamefont {Shekhar}},\ }\href {\doibase
  10.1038/s41467-017-01758-z} {\bibfield  {journal} {\bibinfo  {journal} {Nat.
  Commun.}\ }\textbf {\bibinfo {volume} {8}},\ \bibinfo {pages} {1642}
  (\bibinfo {year} {2017})}\BibitemShut {NoStop}%
\bibitem [{\citenamefont {Wang}\ \emph {et~al.}(2017)\citenamefont {Wang},
  \citenamefont {Graf}, \citenamefont {Liu}, \citenamefont {Du}, \citenamefont
  {Zheng}, \citenamefont {Lei},\ and\ \citenamefont
  {Petrovic}}]{PhysRevB.96.121107}%
  \BibitemOpen
  \bibfield  {author} {\bibinfo {author} {\bibfnamefont {A.}~\bibnamefont
  {Wang}}, \bibinfo {author} {\bibfnamefont {D.}~\bibnamefont {Graf}}, \bibinfo
  {author} {\bibfnamefont {Y.}~\bibnamefont {Liu}}, \bibinfo {author}
  {\bibfnamefont {Q.}~\bibnamefont {Du}}, \bibinfo {author} {\bibfnamefont
  {J.}~\bibnamefont {Zheng}}, \bibinfo {author} {\bibfnamefont
  {H.}~\bibnamefont {Lei}}, \ and\ \bibinfo {author} {\bibfnamefont
  {C.}~\bibnamefont {Petrovic}},\ }\href {\doibase 10.1103/PhysRevB.96.121107}
  {\bibfield  {journal} {\bibinfo  {journal} {Phys. Rev. B}\ }\textbf {\bibinfo
  {volume} {96}},\ \bibinfo {pages} {121107} (\bibinfo {year}
  {2017})}\BibitemShut {NoStop}%
\bibitem [{\citenamefont {Sch\"onemann}\ \emph {et~al.}(2017)\citenamefont
  {Sch\"onemann}, \citenamefont {Aryal}, \citenamefont {Zhou}, \citenamefont
  {Chiu}, \citenamefont {Chen}, \citenamefont {Martin}, \citenamefont
  {McCandless}, \citenamefont {Chan}, \citenamefont {Manousakis},\ and\
  \citenamefont {Balicas}}]{PhysRevB.96.121108}%
  \BibitemOpen
  \bibfield  {author} {\bibinfo {author} {\bibfnamefont {R.}~\bibnamefont
  {Sch\"onemann}}, \bibinfo {author} {\bibfnamefont {N.}~\bibnamefont {Aryal}},
  \bibinfo {author} {\bibfnamefont {Q.}~\bibnamefont {Zhou}}, \bibinfo {author}
  {\bibfnamefont {Y.-C.}\ \bibnamefont {Chiu}}, \bibinfo {author}
  {\bibfnamefont {K.-W.}\ \bibnamefont {Chen}}, \bibinfo {author}
  {\bibfnamefont {T.~J.}\ \bibnamefont {Martin}}, \bibinfo {author}
  {\bibfnamefont {G.~T.}\ \bibnamefont {McCandless}}, \bibinfo {author}
  {\bibfnamefont {J.~Y.}\ \bibnamefont {Chan}}, \bibinfo {author}
  {\bibfnamefont {E.}~\bibnamefont {Manousakis}}, \ and\ \bibinfo {author}
  {\bibfnamefont {L.}~\bibnamefont {Balicas}},\ }\href {\doibase
  10.1103/PhysRevB.96.121108} {\bibfield  {journal} {\bibinfo  {journal} {Phys.
  Rev. B}\ }\textbf {\bibinfo {volume} {96}},\ \bibinfo {pages} {121108}
  (\bibinfo {year} {2017})}\BibitemShut {NoStop}%
\bibitem [{\citenamefont {Lou}\ \emph {et~al.}(2017)\citenamefont {Lou},
  \citenamefont {Fu}, \citenamefont {Xu}, \citenamefont {Guo}, \citenamefont
  {Kong}, \citenamefont {Zeng}, \citenamefont {Ma}, \citenamefont {Richard},
  \citenamefont {Fang}, \citenamefont {Huang}, \citenamefont {Sun},
  \citenamefont {Wang}, \citenamefont {Wang}, \citenamefont {Shi},
  \citenamefont {Lei}, \citenamefont {Liu}, \citenamefont {Weng}, \citenamefont
  {Qian}, \citenamefont {Ding},\ and\ \citenamefont
  {Wang}}]{PhysRevB.95.115140}%
  \BibitemOpen
  \bibfield  {author} {\bibinfo {author} {\bibfnamefont {R.}~\bibnamefont
  {Lou}}, \bibinfo {author} {\bibfnamefont {B.-B.}\ \bibnamefont {Fu}},
  \bibinfo {author} {\bibfnamefont {Q.~N.}\ \bibnamefont {Xu}}, \bibinfo
  {author} {\bibfnamefont {P.-J.}\ \bibnamefont {Guo}}, \bibinfo {author}
  {\bibfnamefont {L.-Y.}\ \bibnamefont {Kong}}, \bibinfo {author}
  {\bibfnamefont {L.-K.}\ \bibnamefont {Zeng}}, \bibinfo {author}
  {\bibfnamefont {J.-Z.}\ \bibnamefont {Ma}}, \bibinfo {author} {\bibfnamefont
  {P.}~\bibnamefont {Richard}}, \bibinfo {author} {\bibfnamefont
  {C.}~\bibnamefont {Fang}}, \bibinfo {author} {\bibfnamefont {Y.-B.}\
  \bibnamefont {Huang}}, \bibinfo {author} {\bibfnamefont {S.-S.}\ \bibnamefont
  {Sun}}, \bibinfo {author} {\bibfnamefont {Q.}~\bibnamefont {Wang}}, \bibinfo
  {author} {\bibfnamefont {L.}~\bibnamefont {Wang}}, \bibinfo {author}
  {\bibfnamefont {Y.-G.}\ \bibnamefont {Shi}}, \bibinfo {author} {\bibfnamefont
  {H.~C.}\ \bibnamefont {Lei}}, \bibinfo {author} {\bibfnamefont
  {K.}~\bibnamefont {Liu}}, \bibinfo {author} {\bibfnamefont {H.~M.}\
  \bibnamefont {Weng}}, \bibinfo {author} {\bibfnamefont {T.}~\bibnamefont
  {Qian}}, \bibinfo {author} {\bibfnamefont {H.}~\bibnamefont {Ding}}, \ and\
  \bibinfo {author} {\bibfnamefont {S.-C.}\ \bibnamefont {Wang}},\ }\href
  {\doibase 10.1103/PhysRevB.95.115140} {\bibfield  {journal} {\bibinfo
  {journal} {Phys. Rev. B}\ }\textbf {\bibinfo {volume} {95}},\ \bibinfo
  {pages} {115140} (\bibinfo {year} {2017})}\BibitemShut {NoStop}%
\bibitem [{\citenamefont {Bl\"ochl}(1994)}]{PhysRevB.50.17953}%
  \BibitemOpen
  \bibfield  {author} {\bibinfo {author} {\bibfnamefont {P.~E.}\ \bibnamefont
  {Bl\"ochl}},\ }\href {\doibase 10.1103/PhysRevB.50.17953} {\bibfield
  {journal} {\bibinfo  {journal} {Phys. Rev. B}\ }\textbf {\bibinfo {volume}
  {50}},\ \bibinfo {pages} {17953} (\bibinfo {year} {1994})}\BibitemShut
  {NoStop}%
\bibitem [{\citenamefont {Kresse}\ and\ \citenamefont
  {Joubert}(1999)}]{PhysRevB.59.1758}%
  \BibitemOpen
  \bibfield  {author} {\bibinfo {author} {\bibfnamefont {G.}~\bibnamefont
  {Kresse}}\ and\ \bibinfo {author} {\bibfnamefont {D.}~\bibnamefont
  {Joubert}},\ }\href {\doibase 10.1103/PhysRevB.59.1758} {\bibfield  {journal}
  {\bibinfo  {journal} {Phys. Rev. B}\ }\textbf {\bibinfo {volume} {59}},\
  \bibinfo {pages} {1758} (\bibinfo {year} {1999})}\BibitemShut {NoStop}%
\bibitem [{\citenamefont {Kresse}\ and\ \citenamefont
  {Hafner}(1993)}]{PhysRevB.47.558}%
  \BibitemOpen
  \bibfield  {author} {\bibinfo {author} {\bibfnamefont {G.}~\bibnamefont
  {Kresse}}\ and\ \bibinfo {author} {\bibfnamefont {J.}~\bibnamefont
  {Hafner}},\ }\href {\doibase 10.1103/PhysRevB.47.558} {\bibfield  {journal}
  {\bibinfo  {journal} {Phys. Rev. B}\ }\textbf {\bibinfo {volume} {47}},\
  \bibinfo {pages} {558} (\bibinfo {year} {1993})}\BibitemShut {NoStop}%
\bibitem [{\citenamefont {Kresse}\ and\ \citenamefont
  {Furthmüller}(1996)}]{KRESSE199615}%
  \BibitemOpen
  \bibfield  {author} {\bibinfo {author} {\bibfnamefont {G.}~\bibnamefont
  {Kresse}}\ and\ \bibinfo {author} {\bibfnamefont {J.}~\bibnamefont
  {Furthmüller}},\ }\href {\doibase
  https://doi.org/10.1016/0927-0256(96)00008-0} {\bibfield  {journal} {\bibinfo
   {journal} {Comp. Mater. Sci.}\ }\textbf {\bibinfo {volume} {6}},\ \bibinfo
  {pages} {15 } (\bibinfo {year} {1996})}\BibitemShut {NoStop}%
\bibitem [{\citenamefont {Kresse}\ and\ \citenamefont
  {Furthm\"uller}(1996)}]{PhysRevB.54.11169}%
  \BibitemOpen
  \bibfield  {author} {\bibinfo {author} {\bibfnamefont {G.}~\bibnamefont
  {Kresse}}\ and\ \bibinfo {author} {\bibfnamefont {J.}~\bibnamefont
  {Furthm\"uller}},\ }\href {\doibase 10.1103/PhysRevB.54.11169} {\bibfield
  {journal} {\bibinfo  {journal} {Phys. Rev. B}\ }\textbf {\bibinfo {volume}
  {54}},\ \bibinfo {pages} {11169} (\bibinfo {year} {1996})}\BibitemShut
  {NoStop}%
\bibitem [{\citenamefont {Perdew}\ \emph {et~al.}(1996)\citenamefont {Perdew},
  \citenamefont {Burke},\ and\ \citenamefont
  {Ernzerhof}}]{PhysRevLett.77.3865}%
  \BibitemOpen
  \bibfield  {author} {\bibinfo {author} {\bibfnamefont {J.~P.}\ \bibnamefont
  {Perdew}}, \bibinfo {author} {\bibfnamefont {K.}~\bibnamefont {Burke}}, \
  and\ \bibinfo {author} {\bibfnamefont {M.}~\bibnamefont {Ernzerhof}},\ }\href
  {\doibase 10.1103/PhysRevLett.77.3865} {\bibfield  {journal} {\bibinfo
  {journal} {Phys. Rev. Lett.}\ }\textbf {\bibinfo {volume} {77}},\ \bibinfo
  {pages} {3865} (\bibinfo {year} {1996})}\BibitemShut {NoStop}%
\bibitem [{\citenamefont {Sun}\ \emph {et~al.}(2015)\citenamefont {Sun},
  \citenamefont {Ruzsinszky},\ and\ \citenamefont
  {Perdew}}]{PhysRevLett.115.036402}%
  \BibitemOpen
  \bibfield  {author} {\bibinfo {author} {\bibfnamefont {J.}~\bibnamefont
  {Sun}}, \bibinfo {author} {\bibfnamefont {A.}~\bibnamefont {Ruzsinszky}}, \
  and\ \bibinfo {author} {\bibfnamefont {J.~P.}\ \bibnamefont {Perdew}},\
  }\href {\doibase 10.1103/PhysRevLett.115.036402} {\bibfield  {journal}
  {\bibinfo  {journal} {Phys. Rev. Lett.}\ }\textbf {\bibinfo {volume} {115}},\
  \bibinfo {pages} {036402} (\bibinfo {year} {2015})}\BibitemShut {NoStop}%
\bibitem [{\citenamefont {Perdew}\ \emph {et~al.}(2005)\citenamefont {Perdew},
  \citenamefont {Ruzsinszky}, \citenamefont {Tao}, \citenamefont {Staroverov},
  \citenamefont {Scuseria},\ and\ \citenamefont
  {Csonka}}]{Perdew_J.Chem.Phys2015}%
  \BibitemOpen
  \bibfield  {author} {\bibinfo {author} {\bibfnamefont {J.~P.}\ \bibnamefont
  {Perdew}}, \bibinfo {author} {\bibfnamefont {A.}~\bibnamefont {Ruzsinszky}},
  \bibinfo {author} {\bibfnamefont {J.}~\bibnamefont {Tao}}, \bibinfo {author}
  {\bibfnamefont {V.~N.}\ \bibnamefont {Staroverov}}, \bibinfo {author}
  {\bibfnamefont {G.~E.}\ \bibnamefont {Scuseria}}, \ and\ \bibinfo {author}
  {\bibfnamefont {G.~I.}\ \bibnamefont {Csonka}},\ }\href {\doibase
  10.1063/1.1904565} {\bibfield  {journal} {\bibinfo  {journal} {J. Chem.
  Phys}\ }\textbf {\bibinfo {volume} {123}},\ \bibinfo {pages} {062201}
  (\bibinfo {year} {2005})}\BibitemShut {NoStop}%
\bibitem [{\citenamefont {Marzari}\ and\ \citenamefont
  {Vanderbilt}(1997)}]{PhysRevB.56.12847}%
  \BibitemOpen
  \bibfield  {author} {\bibinfo {author} {\bibfnamefont {N.}~\bibnamefont
  {Marzari}}\ and\ \bibinfo {author} {\bibfnamefont {D.}~\bibnamefont
  {Vanderbilt}},\ }\href {\doibase 10.1103/PhysRevB.56.12847} {\bibfield
  {journal} {\bibinfo  {journal} {Phys. Rev. B}\ }\textbf {\bibinfo {volume}
  {56}},\ \bibinfo {pages} {12847} (\bibinfo {year} {1997})}\BibitemShut
  {NoStop}%
\bibitem [{\citenamefont {Souza}\ \emph {et~al.}(2001)\citenamefont {Souza},
  \citenamefont {Marzari},\ and\ \citenamefont
  {Vanderbilt}}]{PhysRevB.65.035109}%
  \BibitemOpen
  \bibfield  {author} {\bibinfo {author} {\bibfnamefont {I.}~\bibnamefont
  {Souza}}, \bibinfo {author} {\bibfnamefont {N.}~\bibnamefont {Marzari}}, \
  and\ \bibinfo {author} {\bibfnamefont {D.}~\bibnamefont {Vanderbilt}},\
  }\href {\doibase 10.1103/PhysRevB.65.035109} {\bibfield  {journal} {\bibinfo
  {journal} {Phys. Rev. B}\ }\textbf {\bibinfo {volume} {65}},\ \bibinfo
  {pages} {035109} (\bibinfo {year} {2001})}\BibitemShut {NoStop}%
\bibitem [{\citenamefont {Mostofi}\ \emph {et~al.}(2008)\citenamefont
  {Mostofi}, \citenamefont {Yates}, \citenamefont {Lee}, \citenamefont {Souza},
  \citenamefont {Vanderbilt},\ and\ \citenamefont {Marzari}}]{MOSTOFI2008685}%
  \BibitemOpen
  \bibfield  {author} {\bibinfo {author} {\bibfnamefont {A.~A.}\ \bibnamefont
  {Mostofi}}, \bibinfo {author} {\bibfnamefont {J.~R.}\ \bibnamefont {Yates}},
  \bibinfo {author} {\bibfnamefont {Y.-S.}\ \bibnamefont {Lee}}, \bibinfo
  {author} {\bibfnamefont {I.}~\bibnamefont {Souza}}, \bibinfo {author}
  {\bibfnamefont {D.}~\bibnamefont {Vanderbilt}}, \ and\ \bibinfo {author}
  {\bibfnamefont {N.}~\bibnamefont {Marzari}},\ }\href
  {http://www.sciencedirect.com/science/article/pii/S0010465507004936}
  {\bibfield  {journal} {\bibinfo  {journal} {Comput. Phys. Commun.}\ }\textbf
  {\bibinfo {volume} {178}},\ \bibinfo {pages} {685 } (\bibinfo {year}
  {2008})}\BibitemShut {NoStop}%
\bibitem [{\citenamefont {Kane}\ and\ \citenamefont
  {Mele}(2005)}]{PhysRevLett.95.146802}%
  \BibitemOpen
  \bibfield  {author} {\bibinfo {author} {\bibfnamefont {C.~L.}\ \bibnamefont
  {Kane}}\ and\ \bibinfo {author} {\bibfnamefont {E.~J.}\ \bibnamefont
  {Mele}},\ }\href {\doibase 10.1103/PhysRevLett.95.146802} {\bibfield
  {journal} {\bibinfo  {journal} {Phys. Rev. Lett.}\ }\textbf {\bibinfo
  {volume} {95}},\ \bibinfo {pages} {146802} (\bibinfo {year}
  {2005})}\BibitemShut {NoStop}%
\bibitem [{\citenamefont {Yu}\ \emph {et~al.}(2011)\citenamefont {Yu},
  \citenamefont {Qi}, \citenamefont {Bernevig}, \citenamefont {Fang},\ and\
  \citenamefont {Dai}}]{PhysRevB.84.075119}%
  \BibitemOpen
  \bibfield  {author} {\bibinfo {author} {\bibfnamefont {R.}~\bibnamefont
  {Yu}}, \bibinfo {author} {\bibfnamefont {X.~L.}\ \bibnamefont {Qi}}, \bibinfo
  {author} {\bibfnamefont {A.}~\bibnamefont {Bernevig}}, \bibinfo {author}
  {\bibfnamefont {Z.}~\bibnamefont {Fang}}, \ and\ \bibinfo {author}
  {\bibfnamefont {X.}~\bibnamefont {Dai}},\ }\href {\doibase
  10.1103/PhysRevB.84.075119} {\bibfield  {journal} {\bibinfo  {journal} {Phys.
  Rev. B}\ }\textbf {\bibinfo {volume} {84}},\ \bibinfo {pages} {075119}
  (\bibinfo {year} {2011})}\BibitemShut {NoStop}%
\bibitem [{\citenamefont {Soluyanov}\ and\ \citenamefont
  {Vanderbilt}(2011)}]{PhysRevB.83.235401}%
  \BibitemOpen
  \bibfield  {author} {\bibinfo {author} {\bibfnamefont {A.~A.}\ \bibnamefont
  {Soluyanov}}\ and\ \bibinfo {author} {\bibfnamefont {D.}~\bibnamefont
  {Vanderbilt}},\ }\href {\doibase 10.1103/PhysRevB.83.235401} {\bibfield
  {journal} {\bibinfo  {journal} {Phys. Rev. B}\ }\textbf {\bibinfo {volume}
  {83}},\ \bibinfo {pages} {235401} (\bibinfo {year} {2011})}\BibitemShut
  {NoStop}%
\bibitem [{\citenamefont {Wu}\ \emph {et~al.}(2018)\citenamefont {Wu},
  \citenamefont {Zhang}, \citenamefont {Song}, \citenamefont {Troyer},\ and\
  \citenamefont {Soluyanov}}]{WU2018405}%
  \BibitemOpen
  \bibfield  {author} {\bibinfo {author} {\bibfnamefont {Q.}~\bibnamefont
  {Wu}}, \bibinfo {author} {\bibfnamefont {S.}~\bibnamefont {Zhang}}, \bibinfo
  {author} {\bibfnamefont {H.-F.}\ \bibnamefont {Song}}, \bibinfo {author}
  {\bibfnamefont {M.}~\bibnamefont {Troyer}}, \ and\ \bibinfo {author}
  {\bibfnamefont {A.~A.}\ \bibnamefont {Soluyanov}},\ }\href
  {http://www.sciencedirect.com/science/article/pii/S0010465517303442}
  {\bibfield  {journal} {\bibinfo  {journal} {Comput. Phys. Commun.}\ }\textbf
  {\bibinfo {volume} {224}},\ \bibinfo {pages} {405 } (\bibinfo {year}
  {2018})}\BibitemShut {NoStop}%
\bibitem [{\citenamefont {Noffsinger}\ \emph {et~al.}(2010)\citenamefont
  {Noffsinger}, \citenamefont {Giustino}, \citenamefont {Malone}, \citenamefont
  {Park}, \citenamefont {Louie},\ and\ \citenamefont {Cohen}}]{epw20102140}%
  \BibitemOpen
  \bibfield  {author} {\bibinfo {author} {\bibfnamefont {J.}~\bibnamefont
  {Noffsinger}}, \bibinfo {author} {\bibfnamefont {F.}~\bibnamefont
  {Giustino}}, \bibinfo {author} {\bibfnamefont {B.~D.}\ \bibnamefont
  {Malone}}, \bibinfo {author} {\bibfnamefont {C.-H.}\ \bibnamefont {Park}},
  \bibinfo {author} {\bibfnamefont {S.~G.}\ \bibnamefont {Louie}}, \ and\
  \bibinfo {author} {\bibfnamefont {M.~L.}\ \bibnamefont {Cohen}},\ }\href
  {\doibase https://doi.org/10.1016/j.cpc.2010.08.027} {\bibfield  {journal}
  {\bibinfo  {journal} {Comput. Phys. Commun.}\ }\textbf {\bibinfo {volume}
  {181}},\ \bibinfo {pages} {2140 } (\bibinfo {year} {2010})}\BibitemShut
  {NoStop}%
\bibitem [{\citenamefont {Giustino}\ \emph {et~al.}(2007)\citenamefont
  {Giustino}, \citenamefont {Cohen},\ and\ \citenamefont
  {Louie}}]{PhysRevB.76.165108}%
  \BibitemOpen
  \bibfield  {author} {\bibinfo {author} {\bibfnamefont {F.}~\bibnamefont
  {Giustino}}, \bibinfo {author} {\bibfnamefont {M.~L.}\ \bibnamefont {Cohen}},
  \ and\ \bibinfo {author} {\bibfnamefont {S.~G.}\ \bibnamefont {Louie}},\
  }\href {\doibase 10.1103/PhysRevB.76.165108} {\bibfield  {journal} {\bibinfo
  {journal} {Phys. Rev. B}\ }\textbf {\bibinfo {volume} {76}},\ \bibinfo
  {pages} {165108} (\bibinfo {year} {2007})}\BibitemShut {NoStop}%
\bibitem [{\citenamefont {Giannozzi}\ \emph {et~al.}(2009)\citenamefont
  {Giannozzi}, \citenamefont {Baroni}, \citenamefont {Bonini}, \citenamefont
  {Calandra}, \citenamefont {Car}, \citenamefont {Cavazzoni}, \citenamefont
  {Ceresoli}, \citenamefont {Chiarotti}, \citenamefont {Cococcioni},
  \citenamefont {Dabo}, \citenamefont {Corso}, \citenamefont {de~Gironcoli},
  \citenamefont {Fabris}, \citenamefont {Fratesi}, \citenamefont {Gebauer},
  \citenamefont {Gerstmann}, \citenamefont {Gougoussis}, \citenamefont
  {Kokalj}, \citenamefont {Lazzeri}, \citenamefont {Martin-Samos},
  \citenamefont {Marzari}, \citenamefont {Mauri}, \citenamefont {Mazzarello},
  \citenamefont {Paolini}, \citenamefont {Pasquarello}, \citenamefont
  {Paulatto}, \citenamefont {Sbraccia}, \citenamefont {Scandolo}, \citenamefont
  {Sclauzero}, \citenamefont {Seitsonen}, \citenamefont {Smogunov},
  \citenamefont {Umari},\ and\ \citenamefont {Wentzcovitch}}]{Giannozzi_2009}%
  \BibitemOpen
  \bibfield  {author} {\bibinfo {author} {\bibfnamefont {P.}~\bibnamefont
  {Giannozzi}}, \bibinfo {author} {\bibfnamefont {S.}~\bibnamefont {Baroni}},
  \bibinfo {author} {\bibfnamefont {N.}~\bibnamefont {Bonini}}, \bibinfo
  {author} {\bibfnamefont {M.}~\bibnamefont {Calandra}}, \bibinfo {author}
  {\bibfnamefont {R.}~\bibnamefont {Car}}, \bibinfo {author} {\bibfnamefont
  {C.}~\bibnamefont {Cavazzoni}}, \bibinfo {author} {\bibfnamefont
  {D.}~\bibnamefont {Ceresoli}}, \bibinfo {author} {\bibfnamefont {G.~L.}\
  \bibnamefont {Chiarotti}}, \bibinfo {author} {\bibfnamefont {M.}~\bibnamefont
  {Cococcioni}}, \bibinfo {author} {\bibfnamefont {I.}~\bibnamefont {Dabo}},
  \bibinfo {author} {\bibfnamefont {A.~D.}\ \bibnamefont {Corso}}, \bibinfo
  {author} {\bibfnamefont {S.}~\bibnamefont {de~Gironcoli}}, \bibinfo {author}
  {\bibfnamefont {S.}~\bibnamefont {Fabris}}, \bibinfo {author} {\bibfnamefont
  {G.}~\bibnamefont {Fratesi}}, \bibinfo {author} {\bibfnamefont
  {R.}~\bibnamefont {Gebauer}}, \bibinfo {author} {\bibfnamefont
  {U.}~\bibnamefont {Gerstmann}}, \bibinfo {author} {\bibfnamefont
  {C.}~\bibnamefont {Gougoussis}}, \bibinfo {author} {\bibfnamefont
  {A.}~\bibnamefont {Kokalj}}, \bibinfo {author} {\bibfnamefont
  {M.}~\bibnamefont {Lazzeri}}, \bibinfo {author} {\bibfnamefont
  {L.}~\bibnamefont {Martin-Samos}}, \bibinfo {author} {\bibfnamefont
  {N.}~\bibnamefont {Marzari}}, \bibinfo {author} {\bibfnamefont
  {F.}~\bibnamefont {Mauri}}, \bibinfo {author} {\bibfnamefont
  {R.}~\bibnamefont {Mazzarello}}, \bibinfo {author} {\bibfnamefont
  {S.}~\bibnamefont {Paolini}}, \bibinfo {author} {\bibfnamefont
  {A.}~\bibnamefont {Pasquarello}}, \bibinfo {author} {\bibfnamefont
  {L.}~\bibnamefont {Paulatto}}, \bibinfo {author} {\bibfnamefont
  {C.}~\bibnamefont {Sbraccia}}, \bibinfo {author} {\bibfnamefont
  {S.}~\bibnamefont {Scandolo}}, \bibinfo {author} {\bibfnamefont
  {G.}~\bibnamefont {Sclauzero}}, \bibinfo {author} {\bibfnamefont {A.~P.}\
  \bibnamefont {Seitsonen}}, \bibinfo {author} {\bibfnamefont {A.}~\bibnamefont
  {Smogunov}}, \bibinfo {author} {\bibfnamefont {P.}~\bibnamefont {Umari}}, \
  and\ \bibinfo {author} {\bibfnamefont {R.~M.}\ \bibnamefont {Wentzcovitch}},\
  }\href {\doibase 10.1088/0953-8984/21/39/395502} {\bibfield  {journal}
  {\bibinfo  {journal} {J. Phys.: Condens. Matter}\ }\textbf {\bibinfo {volume}
  {21}},\ \bibinfo {pages} {395502} (\bibinfo {year} {2009})}\BibitemShut
  {NoStop}%
\bibitem [{\citenamefont {Ponc\'e}\ \emph {et~al.}(2018)\citenamefont
  {Ponc\'e}, \citenamefont {Margine},\ and\ \citenamefont
  {Giustino}}]{PhysRevB.97.121201}%
  \BibitemOpen
  \bibfield  {author} {\bibinfo {author} {\bibfnamefont {S.}~\bibnamefont
  {Ponc\'e}}, \bibinfo {author} {\bibfnamefont {E.~R.}\ \bibnamefont
  {Margine}}, \ and\ \bibinfo {author} {\bibfnamefont {F.}~\bibnamefont
  {Giustino}},\ }\href {\doibase 10.1103/PhysRevB.97.121201} {\bibfield
  {journal} {\bibinfo  {journal} {Phys. Rev. B}\ }\textbf {\bibinfo {volume}
  {97}},\ \bibinfo {pages} {121201} (\bibinfo {year} {2018})}\BibitemShut
  {NoStop}%
\bibitem [{\citenamefont {Zhou}\ \emph {et~al.}(2015)\citenamefont {Zhou},
  \citenamefont {Liao}, \citenamefont {Qiu}, \citenamefont {Huberman},
  \citenamefont {Esfarjani}, \citenamefont {Dresselhaus},\ and\ \citenamefont
  {Chen}}]{2015Zhou14777}%
  \BibitemOpen
  \bibfield  {author} {\bibinfo {author} {\bibfnamefont {J.}~\bibnamefont
  {Zhou}}, \bibinfo {author} {\bibfnamefont {B.}~\bibnamefont {Liao}}, \bibinfo
  {author} {\bibfnamefont {B.}~\bibnamefont {Qiu}}, \bibinfo {author}
  {\bibfnamefont {S.}~\bibnamefont {Huberman}}, \bibinfo {author}
  {\bibfnamefont {K.}~\bibnamefont {Esfarjani}}, \bibinfo {author}
  {\bibfnamefont {M.~S.}\ \bibnamefont {Dresselhaus}}, \ and\ \bibinfo {author}
  {\bibfnamefont {G.}~\bibnamefont {Chen}},\ }\href {\doibase
  10.1073/pnas.1512328112} {\bibfield  {journal} {\bibinfo  {journal} {Proc.
  Natl. Acad. Sci}\ }\textbf {\bibinfo {volume} {112}},\ \bibinfo {pages}
  {14777} (\bibinfo {year} {2015})}\BibitemShut {NoStop}%
\bibitem [{\citenamefont {Iandelli}\ and\ \citenamefont
  {Palenzona}(1974)}]{IANDELLI19741}%
  \BibitemOpen
  \bibfield  {author} {\bibinfo {author} {\bibfnamefont {A.}~\bibnamefont
  {Iandelli}}\ and\ \bibinfo {author} {\bibfnamefont {A.}~\bibnamefont
  {Palenzona}},\ }\href
  {http://www.sciencedirect.com/science/article/pii/0022508874901970}
  {\bibfield  {journal} {\bibinfo  {journal} {J. Less-Common Met.}\ }\textbf
  {\bibinfo {volume} {38}},\ \bibinfo {pages} {1 } (\bibinfo {year}
  {1974})}\BibitemShut {NoStop}%
\bibitem [{\citenamefont {N.~W.~Ashcroft}(1976)}]{solid_phys}%
  \BibitemOpen
  \bibfield  {author} {\bibinfo {author} {\bibfnamefont {N.~D.~Mermin}\ and
  \bibnamefont {N.~W.~Ashcroft}},\ }\href@noop {} {\emph {\bibinfo {title}
  {Solid State Physics}}}\ (\bibinfo  {publisher} {Saunders College, New
  York},\ \bibinfo {year} {1976})\BibitemShut {NoStop}%
\bibitem [{\citenamefont {Ziman}(1960)}]{Elec_Phon}%
  \BibitemOpen
  \bibfield  {author} {\bibinfo {author} {\bibfnamefont {J.~M.}\ \bibnamefont
  {Ziman}},\ }\href@noop {} {\emph {\bibinfo {title} {Electrons and Phonons}}}\
  (\bibinfo  {publisher} {Clarendon, Oxford},\ \bibinfo {year}
  {1960})\BibitemShut {NoStop}%
\bibitem [{\citenamefont {Fu}\ \emph {et~al.}(2007)\citenamefont {Fu},
  \citenamefont {Kane},\ and\ \citenamefont {Mele}}]{PhysRevLett.98.106803}%
  \BibitemOpen
  \bibfield  {author} {\bibinfo {author} {\bibfnamefont {L.}~\bibnamefont
  {Fu}}, \bibinfo {author} {\bibfnamefont {C.~L.}\ \bibnamefont {Kane}}, \ and\
  \bibinfo {author} {\bibfnamefont {E.~J.}\ \bibnamefont {Mele}},\ }\href
  {\doibase 10.1103/PhysRevLett.98.106803} {\bibfield  {journal} {\bibinfo
  {journal} {Phys. Rev. Lett.}\ }\textbf {\bibinfo {volume} {98}},\ \bibinfo
  {pages} {106803} (\bibinfo {year} {2007})}\BibitemShut {NoStop}%
\bibitem [{\citenamefont {Zhang}\ \emph {et~al.}(2019)\citenamefont {Zhang},
  \citenamefont {Jiang}, \citenamefont {Song}, \citenamefont {Huang},
  \citenamefont {He}, \citenamefont {Fang}, \citenamefont {Weng},\ and\
  \citenamefont {Fang}}]{2019Cata-topMater1}%
  \BibitemOpen
  \bibfield  {author} {\bibinfo {author} {\bibfnamefont {T.}~\bibnamefont
  {Zhang}}, \bibinfo {author} {\bibfnamefont {Y.}~\bibnamefont {Jiang}},
  \bibinfo {author} {\bibfnamefont {Z.}~\bibnamefont {Song}}, \bibinfo {author}
  {\bibfnamefont {H.}~\bibnamefont {Huang}}, \bibinfo {author} {\bibfnamefont
  {Y.}~\bibnamefont {He}}, \bibinfo {author} {\bibfnamefont {Z.}~\bibnamefont
  {Fang}}, \bibinfo {author} {\bibfnamefont {H.}~\bibnamefont {Weng}}, \ and\
  \bibinfo {author} {\bibfnamefont {C.}~\bibnamefont {Fang}},\ }\href {\doibase
  10.1038/s41586-019-0944-6} {\bibfield  {journal} {\bibinfo  {journal}
  {Nature}\ }\textbf {\bibinfo {volume} {566}},\ \bibinfo {pages} {475}
  (\bibinfo {year} {2019})}\BibitemShut {NoStop}%
\bibitem [{\citenamefont {Vergniory}\ \emph {et~al.}(2019)\citenamefont
  {Vergniory}, \citenamefont {Elcoro}, \citenamefont {Felser}, \citenamefont
  {Regnault}, \citenamefont {Bernevig},\ and\ \citenamefont
  {Wang}}]{2019Cata-topMater2}%
  \BibitemOpen
  \bibfield  {author} {\bibinfo {author} {\bibfnamefont {M.~G.}\ \bibnamefont
  {Vergniory}}, \bibinfo {author} {\bibfnamefont {L.}~\bibnamefont {Elcoro}},
  \bibinfo {author} {\bibfnamefont {C.}~\bibnamefont {Felser}}, \bibinfo
  {author} {\bibfnamefont {N.}~\bibnamefont {Regnault}}, \bibinfo {author}
  {\bibfnamefont {B.~A.}\ \bibnamefont {Bernevig}}, \ and\ \bibinfo {author}
  {\bibfnamefont {Z.}~\bibnamefont {Wang}},\ }\href {\doibase
  10.1038/s41586-019-0954-4} {\bibfield  {journal} {\bibinfo  {journal}
  {Nature}\ }\textbf {\bibinfo {volume} {566}},\ \bibinfo {pages} {480}
  (\bibinfo {year} {2019})}\BibitemShut {NoStop}%
\bibitem [{\citenamefont {Tang}\ \emph {et~al.}(2019)\citenamefont {Tang},
  \citenamefont {Po}, \citenamefont {Vishwanath},\ and\ \citenamefont
  {Wan}}]{2019Cata-topMater3}%
  \BibitemOpen
  \bibfield  {author} {\bibinfo {author} {\bibfnamefont {F.}~\bibnamefont
  {Tang}}, \bibinfo {author} {\bibfnamefont {H.~C.}\ \bibnamefont {Po}},
  \bibinfo {author} {\bibfnamefont {A.}~\bibnamefont {Vishwanath}}, \ and\
  \bibinfo {author} {\bibfnamefont {X.}~\bibnamefont {Wan}},\ }\href {\doibase
  10.1038/s41586-019-0937-5} {\bibfield  {journal} {\bibinfo  {journal}
  {Nature}\ }\textbf {\bibinfo {volume} {566}},\ \bibinfo {pages} {486}
  (\bibinfo {year} {2019})}\BibitemShut {NoStop}%
\bibitem [{\citenamefont {Bardeen}\ and\ \citenamefont
  {Shockley}(1950)}]{PhysRev.80.72}%
  \BibitemOpen
  \bibfield  {author} {\bibinfo {author} {\bibfnamefont {J.}~\bibnamefont
  {Bardeen}}\ and\ \bibinfo {author} {\bibfnamefont {W.}~\bibnamefont
  {Shockley}},\ }\href {\doibase 10.1103/PhysRev.80.72} {\bibfield  {journal}
  {\bibinfo  {journal} {Phys. Rev.}\ }\textbf {\bibinfo {volume} {80}},\
  \bibinfo {pages} {72} (\bibinfo {year} {1950})}\BibitemShut {NoStop}%
\bibitem [{\citenamefont {Xi}\ \emph {et~al.}(2012)\citenamefont {Xi},
  \citenamefont {Long}, \citenamefont {Tang}, \citenamefont {Wang},\ and\
  \citenamefont {Shuai}}]{C2NR30585B}%
  \BibitemOpen
  \bibfield  {author} {\bibinfo {author} {\bibfnamefont {J.}~\bibnamefont
  {Xi}}, \bibinfo {author} {\bibfnamefont {M.}~\bibnamefont {Long}}, \bibinfo
  {author} {\bibfnamefont {L.}~\bibnamefont {Tang}}, \bibinfo {author}
  {\bibfnamefont {D.}~\bibnamefont {Wang}}, \ and\ \bibinfo {author}
  {\bibfnamefont {Z.}~\bibnamefont {Shuai}},\ }\href {\doibase
  10.1039/C2NR30585B} {\bibfield  {journal} {\bibinfo  {journal} {Nanoscale}\
  }\textbf {\bibinfo {volume} {4}},\ \bibinfo {pages} {4348} (\bibinfo {year}
  {2012})}\BibitemShut {NoStop}%
\bibitem [{\citenamefont {Fiori}\ and\ \citenamefont
  {Iannaccone}(2013)}]{Fiori.2013}%
  \BibitemOpen
  \bibfield  {author} {\bibinfo {author} {\bibfnamefont {G.}~\bibnamefont
  {Fiori}}\ and\ \bibinfo {author} {\bibfnamefont {G.}~\bibnamefont
  {Iannaccone}},\ }\href {\doibase 10.1109/JPROC.2013.2259451} {\bibfield
  {journal} {\bibinfo  {journal} {Proc. IEEE}\ }\textbf {\bibinfo {volume}
  {101}},\ \bibinfo {pages} {1653} (\bibinfo {year} {2013})}\BibitemShut
  {NoStop}%
\bibitem [{\citenamefont {Qiao}\ \emph {et~al.}(2014)\citenamefont {Qiao},
  \citenamefont {Kong}, \citenamefont {Hu}, \citenamefont {Yang},\ and\
  \citenamefont {Ji}}]{qiao.nc2014}%
  \BibitemOpen
  \bibfield  {author} {\bibinfo {author} {\bibfnamefont {J.}~\bibnamefont
  {Qiao}}, \bibinfo {author} {\bibfnamefont {X.}~\bibnamefont {Kong}}, \bibinfo
  {author} {\bibfnamefont {Z.-X.}\ \bibnamefont {Hu}}, \bibinfo {author}
  {\bibfnamefont {F.}~\bibnamefont {Yang}}, \ and\ \bibinfo {author}
  {\bibfnamefont {W.}~\bibnamefont {Ji}},\ }\href {\doibase 10.1038/ncomms5475}
  {\bibfield  {journal} {\bibinfo  {journal} {Nat. Commun}\ }\textbf {\bibinfo
  {volume} {5}},\ \bibinfo {pages} {4475} (\bibinfo {year} {2014})}\BibitemShut
  {NoStop}%
\bibitem [{\citenamefont {Zhang}\ \emph {et~al.}(2014)\citenamefont {Zhang},
  \citenamefont {Huang}, \citenamefont {Zhang},\ and\ \citenamefont
  {Li}}]{Zhang2014}%
  \BibitemOpen
  \bibfield  {author} {\bibinfo {author} {\bibfnamefont {W.}~\bibnamefont
  {Zhang}}, \bibinfo {author} {\bibfnamefont {Z.}~\bibnamefont {Huang}},
  \bibinfo {author} {\bibfnamefont {W.}~\bibnamefont {Zhang}}, \ and\ \bibinfo
  {author} {\bibfnamefont {Y.}~\bibnamefont {Li}},\ }\href {\doibase
  10.1007/s12274-014-0532-x} {\bibfield  {journal} {\bibinfo  {journal} {Nano
  Res.}\ }\textbf {\bibinfo {volume} {7}},\ \bibinfo {pages} {1731} (\bibinfo
  {year} {2014})}\BibitemShut {NoStop}%
\bibitem [{\citenamefont {Larson}(2006)}]{PhysRevB.74.205113}%
  \BibitemOpen
  \bibfield  {author} {\bibinfo {author} {\bibfnamefont {P.}~\bibnamefont
  {Larson}},\ }\href {\doibase 10.1103/PhysRevB.74.205113} {\bibfield
  {journal} {\bibinfo  {journal} {Phys. Rev. B}\ }\textbf {\bibinfo {volume}
  {74}},\ \bibinfo {pages} {205113} (\bibinfo {year} {2006})}\BibitemShut
  {NoStop}%
\bibitem [{\citenamefont {Wang}\ and\ \citenamefont
  {Cagin}(2007)}]{PhysRevB.76.075201}%
  \BibitemOpen
  \bibfield  {author} {\bibinfo {author} {\bibfnamefont {G.}~\bibnamefont
  {Wang}}\ and\ \bibinfo {author} {\bibfnamefont {T.}~\bibnamefont {Cagin}},\
  }\href {\doibase 10.1103/PhysRevB.76.075201} {\bibfield  {journal} {\bibinfo
  {journal} {Phys. Rev. B}\ }\textbf {\bibinfo {volume} {76}},\ \bibinfo
  {pages} {075201} (\bibinfo {year} {2007})}\BibitemShut {NoStop}%
\bibitem [{\citenamefont {Morikawa}\ \emph {et~al.}(2018)\citenamefont
  {Morikawa}, \citenamefont {Inamoto},\ and\ \citenamefont
  {Takashiri}}]{Nanotechnology-29-075701}%
  \BibitemOpen
  \bibfield  {author} {\bibinfo {author} {\bibfnamefont {S.}~\bibnamefont
  {Morikawa}}, \bibinfo {author} {\bibfnamefont {T.}~\bibnamefont {Inamoto}}, \
  and\ \bibinfo {author} {\bibfnamefont {M.}~\bibnamefont {Takashiri}},\ }\href
  {http://stacks.iop.org/0957-4484/29/i=7/a=075701} {\bibfield  {journal}
  {\bibinfo  {journal} {Nanotech}\ }\textbf {\bibinfo {volume} {29}},\ \bibinfo
  {pages} {075701} (\bibinfo {year} {2018})}\BibitemShut {NoStop}%
\end{thebibliography}

%

\end{document}